\documentclass[twocolumn]{IEEEtran}
\usepackage[T1]{fontenc}
\usepackage[latin9]{inputenc}
\usepackage{color}
\usepackage{verbatim}
\usepackage{float}
\usepackage{amsmath}
\usepackage{amsthm}
\usepackage{amssymb}
\usepackage{graphicx}
\usepackage{wasysym}
\usepackage{esint}
\usepackage[unicode=true,
 bookmarks=true,bookmarksnumbered=true,bookmarksopen=true,bookmarksopenlevel=1,
 breaklinks=false,pdfborder={0 0 0},pdfborderstyle={},backref=false,colorlinks=false]
 {hyperref}
\hypersetup{pdftitle={Your Title},
 pdfauthor={Your Name},
 pdfpagelayout=OneColumn, pdfnewwindow=true, pdfstartview=XYZ, plainpages=false}

\makeatletter

\providecommand{\tabularnewline}{\\}
\floatstyle{ruled}
\newfloat{algorithm}{tbp}{loa}
\providecommand{\algorithmname}{Algorithm}
\floatname{algorithm}{\protect\algorithmname}

\let\oldforeign@language\foreign@language
\DeclareRobustCommand{\foreign@language}[1]{%
  \lowercase{\oldforeign@language{#1}}}
\theoremstyle{plain}

\theoremstyle{plain}
\usepackage{amsmath, amsthm}
\newtheorem{lem}{Lemma}
\theoremstyle{remark}
\newtheorem{claim}{Claim}

\usepackage[caption=false,font=footnotesize]{subfig}
\usepackage{amsmath}
\usepackage{cuted}

\usepackage{algpseudocode}

\@ifundefined{showcaptionsetup}{}{%
 \PassOptionsToPackage{caption=false}{subfig}}
\usepackage{subfig}
\makeatother

\providecommand{\theoremname}{Theorem}

\begin{document}
\title{Joint Channel Estimation and Cooperative Localization for Near-Field
Ultra-Massive MIMO}
\author{Ruoxiao~Cao,~\IEEEmembership{Graduate~Student~Member,~IEEE,}~Hengtao~He,~\IEEEmembership{Member,~IEEE,}\\
Xianghao~Yu,~\IEEEmembership{Senior Member,~IEEE,}~Shenghui~Song,~\IEEEmembership{Senior~Member,~IEEE,}~Kaibin~Huang,~\IEEEmembership{Fellow,~IEEE,}\\~Jun~Zhang,~\IEEEmembership{Fellow,~IEEE,}~Yi~Gong,~\IEEEmembership{Senior~Member,~IEEE,}~Khaled~B.~Letaief,~\IEEEmembership{Fellow,~IEEE}\vspace{-1em}\thanks{Ruoxiao~Cao is with the Department of Electronic and Computer
Engineering, The Hong Kong University of Science and Technology, Hong
Kong SAR, and also with the Department of Electrical and Electronics
Engineering, Southern University of Science and Technology, Shenzhen
518055, China. (e-mail: \protect\href{mailto:rcaoah@connect.ust.hk}{rcaoah@connect.ust.hk}).}\thanks{Hengtao He is with the School of Information Science and Engineering,
Southeast University, Nanjing, China. (e-mail: \protect\href{mailto:hehengtao@seu.edu.cn}{hehengtao@seu.edu.cn}).}\thanks{Shenghui~Song, Jun~Zhang, and Khaled~B.~Letaief are with the the
Department of Electronic and Computer Engineering, The Hong Kong University
of Science and Technology, Hong Kong SAR. (e-mail: \protect\href{mailto:eeshsong@ust.hk}{eeshsong@ust.hk};
\protect\href{mailto:eejzhang@ust.hk}{eejzhang@ust.hk}; \protect\href{mailto:eekhaled@ust.hk}{eekhaled@ust.hk}).}\thanks{Xianghao~Yu is with the Department of Electrical Engineering, City
University of Hong Kong, Hong Kong SAR. (e-mail: \protect\href{mailto:alex.yu@cityu.edu.hk}{alex.yu@cityu.edu.hk}).}\thanks{Kaibin Huang is with the Department of Electrical and Electronic Engineering,
University of Hong Kong, Hong Kong SAR. (e-mail: \protect\href{mailto:huangkb@eee.hku.hk}{huangkb@eee.hku.hk}).}\thanks{Yi~Gong is with the Department of Electrical and Electronics Engineering,
Southern University of Science and Technology, Shenzhen 518055, China
(e-mail: \protect\href{mailto:gongy@sustech.edu.cn}{gongy@sustech.edu.cn}).}}
\markboth{}{}
\maketitle
\begin{abstract}
The next-generation wireless networks are envisioned to jointly support
high-rate communications and ubiquitous sensing. \textcolor{black}{Ultra-Massive
Multiple-Input Multiple-Output (UM-MIMO)} offers abundant spatial
\textcolor{black}{Degrees of Freedom (DoFs)} for both functions, yet
its large aperture shifts electromagnetic propagation into the near
field, invalidating conventional far-field (plane-wave) assumptions.
While near-field channel modeling has been studied, existing channel
estimation methods are inadequate: on-grid designs suffer from non-orthogonal
codebooks, and off-grid methods lack convergence guarantees, yielding
unreliable estimates. Moreover, channel estimation and localization
are typically designed in isolation, preventing the exchange of information
that could otherwise enable mutual performance improvement. To address
this difficulty, we propose a unified framework that exploits near-field
characteristics to jointly design channel estimation and cooperative
localization. Specifically, we develop a \textcolor{black}{Variational
Newtonized Near-field Channel Estimation (VNNCE)} algorithm that extracts
position-aware soft information from the channel, and a \textcolor{black}{Gaussian
Fusion Cooperative Localization (GFCL)} method that leverages this
information across multiple \textcolor{black}{Base Stations (BSs)}
for enhanced accuracy. The two components are tightly integrated into
a joint architecture that enables localization-aided channel refinement
and, conversely, channel-informed localization\textemdash yielding
mutual performance gains. Simulation results demonstrate that the
proposed VNNCE algorithm outperforms the state-of-the-art baseline,
achieving channel estimation accuracy that approaches the \textcolor{black}{Cram\'{e}r\textendash Rao
Lower Bound (CRLB)}. Furthermore, GFCL \textcolor{black}{achieves highly
accurate and robust localization performance, outperforming conventional baselines in the considered simulations}. 
\end{abstract}


\IEEEpeerreviewmaketitle{}

\section{Introduction}

The next-generation wireless networks, i.e., 6G, are expected to enable
various location-based services and applications, such as Internet
of Things (IoT) \cite{nguyen20216g}, \textcolor{black}{Unmanned Aerial
Vehicles (UAVs)} \cite{zeng2019accessing}, and \textcolor{black}{Vehicle-to-Everything
(V2X)} \cite{luettel2012autonomous}. Achieving high-accuracy localization
over wireless networks will thus be a pivotal enabler for 6G \cite{rappaport2019wireless}.
Traditionally, localization has been designed as a standalone functionality,
requiring dedicated spectral resources and incurring substantial hardware
overhead. To improve the efficiency, a growing trend is to tightly
integrate communication and localization capabilities within a unified
infrastructure, thereby improving spectral efficiency and reducing
hardware costs \cite{liu2022survey,xie2023collaborative,xie2025sensing}.
In particular, the joint design of channel estimation and localization
systems, by sharing frequency band and hardware, is currently attracting
extensive research interests. 

\textcolor{black}{Ultra-Massive Multiple-Input Multiple-Output (UM-MIMO)}
is a key enabler of this vision, offering abundant spatial \textcolor{black}{Degrees
of Freedom (DoFs)} for both communication and localization. However,
the large aperture of UM-MIMO arrays fundamentally alters the electromagnetic
propagation environment: it shifts typical user deployments from the
far field into the near field \cite{chataut2020massive,cao2024newtonized,zheng2025convolutional}.
While equipping \textcolor{black}{Base Stations (BSs)} with hundreds
or even thousands of antennas delivers unprecedented spectral and
energy efficiency \,\cite{yu2023adaptive}, it simultaneously invalidates
the planar wavefront assumption underlying conventional channel estimation
and localization designs. In this near-field regime, accurate modeling
necessitates the use of the spherical wave model \cite{yu2023blind}.
Although this model offers richer spatial DoFs, its highly nonlinear
structure poses significant challenges for both channel estimation
and localization. These considerations underscore the need for novel
signal processing frameworks specifically tailored to near-field UM-MIMO
systems. 

Existing near-field channel estimation approaches can typically be
categorized into on-grid and off-grid methods. The on-grid methods
\cite{cui2022channel} focus on designing a near-field codebook to
incorporate both angle and distance parameters. However, the inherent
difficulty in achieving orthogonality among codewords limits their
estimation accuracy. Although not requiring codebook orthogonality,
existing near-field off-grid methods, including the line search method
\cite{cui2022channel} and the Newton\textquoteright s method \cite{lu2022near},
struggle with achieving satisfactory results. In particular, the lack
of convergence guarantees in these off-grid methods is especially
problematic and \text{blue}{can limit their estimation accuracy} for
high-precision localization systems.

Recent advances in localization have increasingly embraced cooperative
frameworks that exploit spatial diversity across multiple \textcolor{black}{BSs}
to improve accuracy. Techniques such as sum-product message passing\,\cite{wymeersch2009cooperative},
\textcolor{black}{Belief Propagation (BP)}\,\cite{meyer2015distributed},
and soft information fusion\,\cite{mazuelas2018soft} enable statistically
principled integration of measurements from distributed nodes. These
methods significantly outperform single-BS approaches by fusing measurements
across multiple BSs, and this advantage is further enhanced in the
dense BS deployments anticipated for 6G.

Despite these advances, a critical gap remains: channel estimation
and cooperative localization for UM-MIMO systems are still treated
as decoupled tasks. Although Yang et al. \cite{yang2022soft} proposed
a soft channel estimation and localization algorithm that leverages
uncertainty-aware information to approach the theoretical performance
limit, their method relies on conventional far-field models. On the
other hand, recent near-field localization approaches \cite{guidi2021radio,guerra2021near,cao2022belief}
operate directly on noisy raw measurements rather than extracting
position-related channel parameters, making them ill-suited for tight
integration with communication signal processing. Moreover, the framework
in \cite{liu2025sensing} performs localization prior to channel estimation
in the near-field regime, but its performance gains are contingent
upon additional hardware support that is not available in standard
UM-MIMO deployments. Consequently, under practical hardware constraints,
most existing work cannot leverage localization feedback to refine
channel estimates, nor can it exploit near-field channel characteristics
to guide position inference. This lack of bidirectional integration
prevents the mutual performance gains that are essential for high-accuracy
6G UM-MIMO systems.

In this work, we address the intertwined challenges of near-field
channel estimation and cooperative localization in UM-MIMO systems
operating under spherical wavefronts. The main contributions are summarized
as follows:
\begin{itemize}
\item We develop a \textcolor{black}{Variational Newtonized Near-field Channel
Estimation (VNNCE)} algorithm that reformulates the nonlinear estimation
problem via variational inference and solves it using a specially
tailored Newton-type optimization algorithm. To ensure convergence
\textcolor{black}{for the individual path refinement stage}, we introduce
a novel near-field codebook that explicitly accounts for the curvature
of spherical wavefronts and meets the requirements for convergence
guarantee of Newton\textquoteright s method. The resulting algorithm
is tuning-free, computationally efficient, and \textcolor{black}{enjoys
local convergence guarantees for the refinement stage,} addressing
a key limitation of existing off-grid methods. 
\item Building on the soft channel estimates from \textcolor{black}{the} VNNCE algorithm, we propose
a \textcolor{black}{Gaussian Fusion Cooperative Localization (GFCL)}
algorithm for multiple BSs networks. GFCL first generates per-BS soft
position estimates and then fuses them in a statistically principled
manner to significantly enhance localization accuracy. Moreover, we
design a joint estimation\textendash localization architecture that
enables bidirectional refinement: localization results feed back to
improve channel estimation, while refined channel estimates further
sharpen position inference. This closed-loop integration unlocks synergistic
gains unattainable in decoupled designs. 
\item Simulation results demonstrate that the VNNCE algorithm achieves channel
estimation accuracy approaching the \textcolor{black}{Cram\'{e}r-Rao Lower
Bound (CRLB)} and outperforms the state-of-the-art baselines. In a
practical near-field UM-MIMO system, the GFCL algorithm enhances localization
accuracy from meter-level to millimeter-level precision. Moreover,
the high-accuracy position estimates obtained from GFCL provide valuable
side information that further refines channel estimation within the
joint architecture, enabling performance to approach fundamental theoretical
limits. 
\end{itemize}
\par The rest of this paper is organized as follows. In Sec. \ref{sec:SystemModel},
we introduce the system model and the problem formulation. In Sec.
\ref{sec:SCE}, the near-field soft channel estimation algorithm,
i.e., the VNNCE algorithm, is \textcolor{black}{presented}. In Sec. \ref{sec:Codebook-Design},
the requirements of the near-field codebook are derived. In Sec. \ref{sec:SCL},
the GFCL algorithm and the joint architecture are proposed. In Sec.
\ref{sec:Simulation-Results}, the simulation results are provided
to illustrate the advantages of our proposed algorithm and architecture.
Finally, we conclude the paper in Sec. \ref{sec:Conclusions}.

\emph{Notations:} For any matrix $\mathbf{A}$, $\mathbf{A}^{T}$,
$\mathbf{A}^{-1}$ and $\mathbf{A}^{H}$ denote the transpose, inverse
and conjugate transpose of $\mathbf{A}$, respectively. $\mathbf{I}_{M}$
represents the identity matrix of size $N\times N$, and $\mathbf{1}_{N}$
and $\mathbf{0}_{N}$ are the $N$-dimensional all-ones and all-zeros
vectors, respectively. A \textcolor{black}{Probability Density Function
(PDF)} is denoted by $p(\cdot)$. \textcolor{black}{$\mathcal{N}(\boldsymbol{\mu},\mathbf{\Sigma})$
and $\mathcal{CN}(\boldsymbol{\mu},\mathbf{\Sigma})$ denote the Gaussian
and the complex Gaussian distribution} with mean vector $\boldsymbol{\mu}$
and covariance matrix $\mathbf{\Sigma}$, while $f_{\mathrm{N}}(\mathbf{x};\boldsymbol{\mu},\mathbf{\Sigma})$
and $f_{\mathrm{CN}}(\mathbf{x};\boldsymbol{\mu},\mathbf{\Sigma})$
represent their \textcolor{black}{PDFs}. The angle of a complex number $x$ is represented
by $\angle x$. $\mathfrak{R}\{\cdot\}$ denotes the real part of
the complex number. $F(\cdot)$ is the Fisher Information Matrix (FIM). 

\section{System Model and Problem Formulation\label{sec:SystemModel}}

In this section, we first introduce the signal model and the spatial
model for UM-MIMO systems. Then, we elaborate the near-field channel
characteristic and formulate the joint channel estimation and cooperative
localization problem.

\subsection{Signal Model }

We investigate an uplink time division duplexing based narrowband
communication system consisting of a single-antenna user and $I$
BSs%
. The carrier frequency is denoted as $f$, and the wavelength is
given by $\lambda=c/f$, where $c$ is the speed of light. We assume
that each BS employs a \textcolor{black}{Uniform Linear Array (ULA)}
with $M$ antennas spaced at half-wavelength intervals, i.e., $d=\lambda/2$.
The channel estimation for each BS is independent, so we consider
an arbitrary BS without loss of generality and omit the superscript
$i$ in signal model. During the uplink training stage, the received
signal $\mathbf{y}\in\mathbb{C}^{M}$ (observed variable or measurement)
is given by
\begin{equation}
\mathbf{y}=\mathbf{h}x+\mathbf{n},\label{eq:SignalModel}
\end{equation}
where $\mathbf{h}\in\mathbb{C}^{M}$ is the channel vector, $x$ is
the known pilot signal, and $\mathbf{n}\in\mathbb{C}^{M}$ is the
additive complex noise. 

\begin{figure*}[t]
\begin{centering}
\subfloat[]{
\centering{}\includegraphics[height=2.5cm]{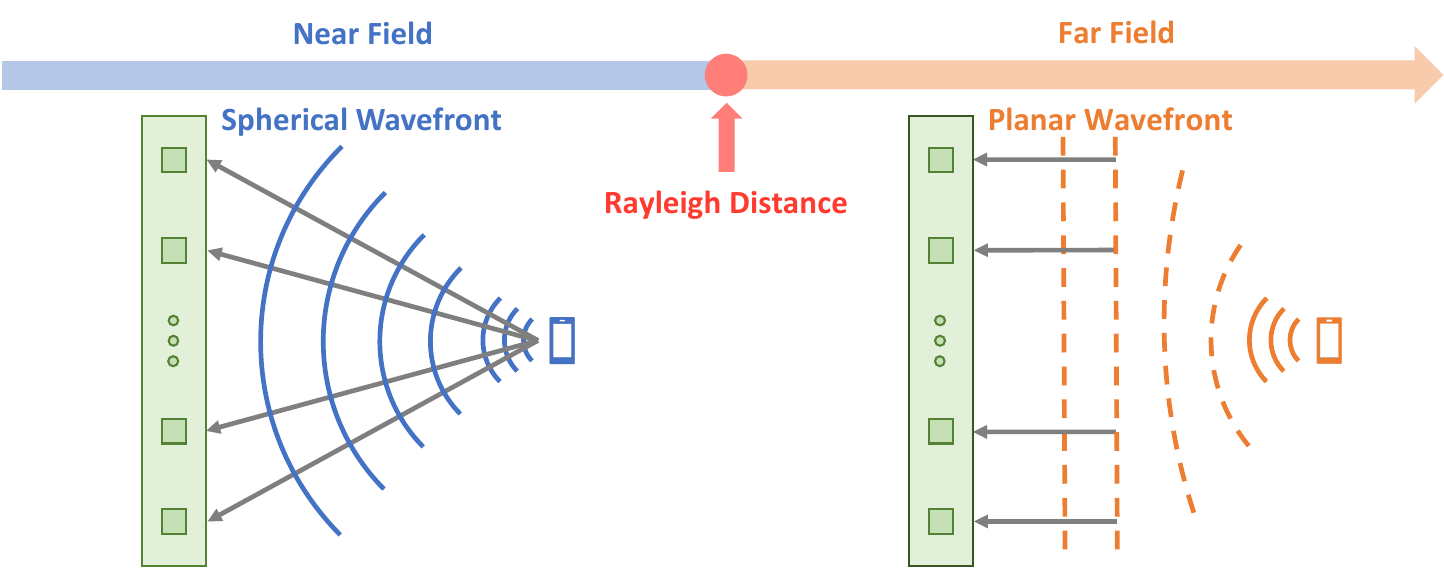}}\quad \quad\subfloat[]{\centering{}\includegraphics[height=3cm]{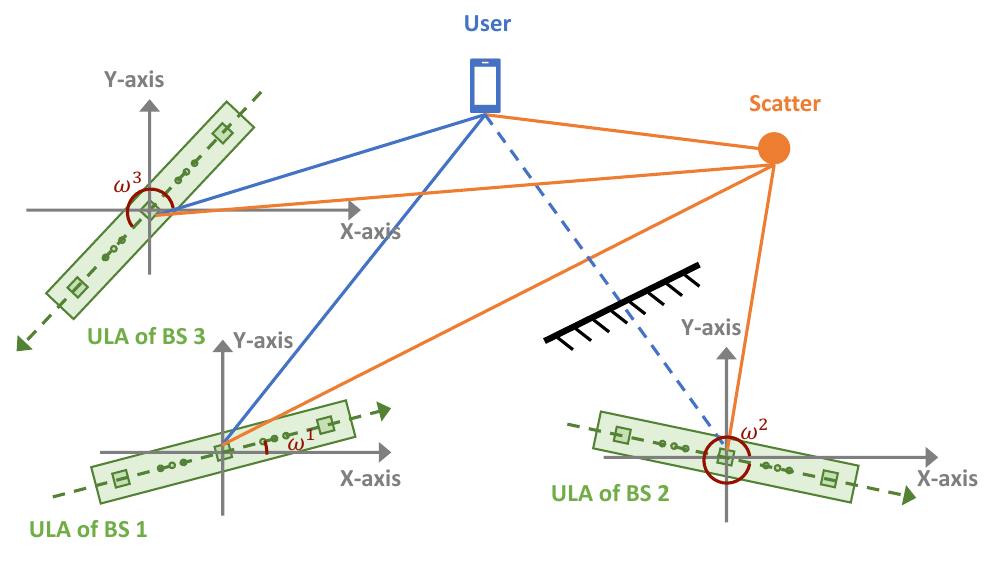}}\quad \quad\subfloat[]{\centering{}\includegraphics[height=3cm]{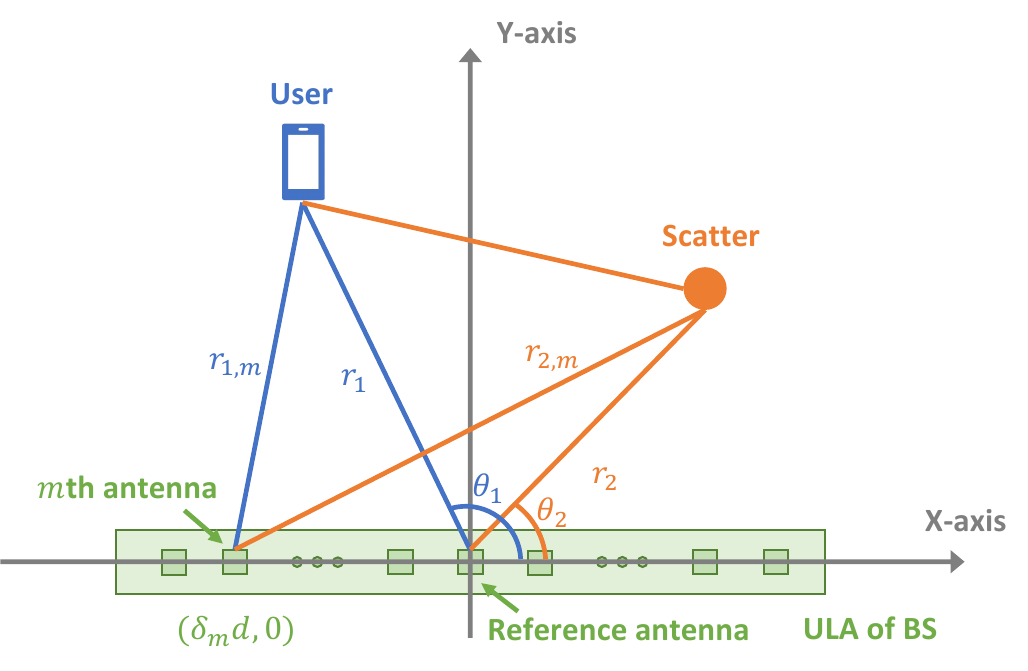}}
\par\end{centering}
\centering{}\caption{(a) The near-field spherical model and far-field planar model. (b)
A sample \textcolor{black}{spatial} model with three BSs. (c) A sample two-path near-field
channel. \label{fig: System-Model}}
\end{figure*}

In conventional MIMO systems, users are assumed to lie in the far-field
region of the BS, where the planar wave assumption holds. However,
the large aperture of UM-MIMO arrays shifts typical user deployments
into the near-field region, invalidating the far-field model. \textcolor{black}{As
shown in Fig. \ref{fig: System-Model}(a), the radiation field is
separated into near-field and far-field regions by the Rayleigh distance
$r_{R}=2D^{2}/\lambda$, where $D$ is the array aperture and $\lambda$
is the carrier wavelength \cite{kraus2002antennas}. For UM-MIMO systems,
where} $M\gg1$, the array aperture of the ULA can be approximately
given by $D\approx Md$, and thus the Rayleigh distance $r_{R}=M^{2}\lambda/2$
is proportional to $M^{2}$. For instance, a 256-element ULA operating
at a carrier frequency of $100\ \textrm{GHz}$ has an aperture of
approximately $0.38\ \textrm{m}$ and a Rayleigh distance of about
$98\ \textrm{m}$\textemdash covering a significant portion of a typical
cellular area. Consequently, the near-field region becomes non-negligible
in future UM-MIMO systems. 

\subsection{Spatial Model}

Each BS $i$, where $i\in\{1,2,\ldots,I\}$ is the index of the BS,
receives uplink wireless signals through its ULA. The channel consists
of one \textcolor{black}{Line-of-Sight (LoS)} path and several \textcolor{black}{Non-Line-of-Sight
(NLoS)} paths. \textcolor{black}{Fig. \ref{fig: System-Model}(b)}
illustrates an example scenario with one user, one scatterer and $I=3$
BSs. We observe that BS 1 receives one LoS path and one NLoS path,
while BS 2 receives only one NLoS path due to the obstruction of the
LoS path. 

Let $\mathbf{x}_{u}=(x_{u},y_{u})^{T}$ denote the position vector
of the user. For BS $i$, the known position vector is $\mathbf{x}_{b}^{i}=(x_{b}^{i},y_{b}^{i})^{T}$,
and $\omega^{i}$ denotes the ULA rotation angle from the positive
X-axis to the ULA orientation. Assuming the LoS path is present, the
user\textquoteright s unknown position can be determined from the
LoS propagation parameters $(\theta_{\textrm{LoS}}^{i},r_{\textrm{LoS}}^{i})$
using the spatial model:

\begin{equation}
\left[\begin{array}{c}
x_{\mathrm{u}}\\
y_{\mathrm{u}}
\end{array}\right]=\left[\begin{array}{c}
x_{\mathrm{b}}^{i}\\
y_{\mathrm{b}}^{i}
\end{array}\right]+r_{\textrm{LoS}}^{i}\left[\begin{array}{c}
\cos\left(\theta_{\textrm{LoS}}^{i}+\omega^{i}\right)\\
\sin\left(\theta_{\textrm{LoS}}^{i}+\omega^{i}\right)
\end{array}\right].\label{eq:SpaceModel}
\end{equation}
Therefore, given estimates $(\hat{\theta}_{\textrm{LoS}}^{i},\hat{r}_{\textrm{LoS}}^{i})$
of the LoS path parameters, the user position estimate $\mathbf{\hat{x}}_{u}=(\hat{x}_{u},\hat{y}_{u})^{T}$
can be obtained. However, it is important to note that the LoS path
may be obstructed by obstacles such as walls. In such cases, if the
NLoS interference cannot be distinguished from the LoS component,
the spatial model in \eqref{eq:SpaceModel} will yield the position
of the scatterer rather than the user, leading to severe degradation
in positioning accuracy. 

\subsection{Problem Formulation}

As illustrated in Fig. \textcolor{black}{\ref{fig: System-Model}}(c),
we consider the near-field channel model given by
\begin{equation}
\mathbf{h}^{\textrm{near}}=\sum_{l=1}^{L}g_{l}e^{j\phi_{l}}\mathbf{b}\left(\theta_{l},r_{l}\right),\label{eq:NCM}
\end{equation}
where $L$ is the number of propagation paths. For each path $l$,
$g_{l}e^{j\phi_{l}}$ denotes the complex path gain at the reference
antenna, $\text{\ensuremath{\theta_{l}}\ensuremath{\ensuremath{\in}(0,\ensuremath{\pi})}}$
is the \textcolor{black}{Angle-of-Arrival (AoA)}, and $r_{l}\in(0,r_{R})$
is the distance between the reference antenna of the ULA and the user
or scatterer. 

Fig. \textcolor{black}{\ref{fig: System-Model}}(c) depicts a two-path
near-field channel, where $r_{1}$ corresponds to the direct path
between the user and the BS, and $r_{2}$ represents the scattered
path between the scatterer and the BS. Note that we neglect the amplitude
variations across array antennas in \eqref{eq:NCM}, as they are insignificant
when $r_{l}>1.2D$ \cite{bjornson2021primer}. 

In contrast to the far-field steering vector based on the planar wave
model, the near-field steering vector $\mathbf{b}(\theta_{l},r_{l})$,
derived from the spherical wave model, is given by 

\begin{equation}
\mathbf{b}\left(\theta_{l},r_{l}\right)=\left[e^{jk_{c}\left(r_{l,1}-r_{l}\right)},\cdots,e^{jk_{c}\left(r_{l,M}-r_{l}\right)}\right]^{T},\label{eq:NSV}
\end{equation}
where $m\in\{1,2,\ldots,M\}$ is the antenna index, $k_{c}=2\pi/\lambda$
is the wavenumber, and $r_{l,m}=\sqrt{r_{l}^{2}+\delta_{m}^{2}d^{2}+2\delta_{m}dr_{l}\cos\theta_{l}}$
denotes the distance between the $m$th antenna and the user or scatterer
of the $l$th path. The coefficient $\delta_{m}=\frac{2m-M+1}{2}$
represents the coordinate of the $m$th antenna. Unlike the planar
wave model in the far field, the spherical wave model implies that
the phase of each element in $\mathbf{b}\left(\theta_{l},r_{l}\right)$
varies nonlinearly with the antenna index $m$. Consequently, the
energy of a near-field path is no longer concentrated at a single
angular direction, and using only an angle parameter is insufficient
to characterize the channel accurately. Thus, for a BS located in
the near field, the received pilot signal in \eqref{eq:SignalModel}
is given by\footnote{We primarily consider a simplified denoising problem for convenience
in this paper. Note that the proposed algorithm can be easily extended
to the problem of compressive near-field channel estimation utilizing
a similar idea from \cite{metzler2016denoising}.} 
\begin{equation}
\mathbf{y}=\mathbf{h}x+\mathbf{n}=\sum_{l=1}^{L}g_{l}e^{j\phi_{l}}\mathbf{b}\left(\theta_{l},r_{l}\right)x+\mathbf{n},\label{eq:NSM}
\end{equation}
where the $L$-path channel $\mathbf{h}$ is determined by unknown
parameters $\boldsymbol{\mu}=\{\theta_{l},r_{l},g_{l},\phi_{l}:l=1,2,\ldots,L\}$,
and the noise \textcolor{black}{$\mathbf{n}\sim\mathcal{CN}\left(\mathbf{0}_{M},\sigma^{2}\mathbf{I}_{M}\right)$}
is assumed to be complex Gaussian noise with \textcolor{black}{covariance
matrix $\sigma^{2}\mathbf{I}_{M}$}. 

The objective of the soft channel estimation is to provide reliable
estimates of unknown parameters $\boldsymbol{\mu}$ along with their
confidence levels, based on the measurement $\mathbf{y}$, for use
in localization. The goal of the soft cooperative localization is
to infer the user's position by fusing this soft information \textcolor{black}{across} multiple
BSs while mitigating interference from NLoS paths. Finally, the refined
position estimate is leveraged to enhance the accuracy of channel
estimation, thereby closing the loop between channel estimation and
localization. 

\section{Near-Field Soft Channel Estimation \label{sec:SCE}}

In this section, we present a novel near-field channel estimation
algorithm VNNCE based on variational inference. The proposed VNNCE
algorithm estimates the key channel parameters\textemdash namely,
angles, distances, and complex gains\textemdash by approximating their
posterior distributions with multivariate Gaussian densities, thereby
yielding soft estimates. Furthermore, we derive the \textcolor{black}{CRLB}
for near-field channel parameter estimation to characterize the fundamental
performance limits of the proposed approach. 

\subsection{Bayesian Formulation }

We first take the estimation of a single path as an example, where
$L=1$ and omit the subscript $l$. The unknown channel parameters
are denoted by $\boldsymbol{\mu}=(\theta,r,g,\phi)^{T}$. We aim to
find the posterior PDF, given by

\begin{equation}
p(\boldsymbol{\mu}\mid\mathbf{y})=\frac{p(\mathbf{y},\boldsymbol{\mu})}{p(\mathbf{y})},\label{eq:posteriorPDF}
\end{equation}
where the denominator $p(\mathbf{y})$, known as the model evidence,
is the marginal probability of the joint PDF and acts as a normalizing
constant. The joint PDF in the numerator \eqref{eq:posteriorPDF}
is the product of the prior PDF and the likelihood function, i.e.,
\begin{equation}
\begin{aligned}p(\mathbf{y},\boldsymbol{\mu}) & =p(\boldsymbol{\mu})p(\mathbf{y}\mid\boldsymbol{\mu}).\end{aligned}
\label{eq:jointPDF}
\end{equation}
Since the elements of the noise $\mathbf{n}$ \textcolor{black}{are}
independent and identically distributed (i.i.d.) complex Gaussian
\textcolor{black}{random variables} with zero mean and variance $\sigma^{2}$,
the likelihood function is given by
\begin{align}
p(\mathbf{y}\mid\boldsymbol{\mu}) & =f_{\mathrm{CN}}(\mathbf{y};ge^{j\phi}\mathbf{b}\left(\theta,r\right),\sigma^{2}\mathbf{I}_{M})\nonumber \\
 & =\prod_{m=1}^{M}f_{\mathrm{CN}}\left(y_{m};ge^{j\phi}e^{jk_{c}\left(r_{m}-r\right)},\sigma^{2}\right),\label{eq:likelihoodY}
\end{align}
where $y_{m}$ represents the $m$th element of vector $\mathbf{y}\in\mathbb{C}^{M}$. 

Unfortunately, obtaining analytical solutions for $\boldsymbol{\mu}$
is intractable. Therefore, we employ a type of approximation technique
called variational inference to enable statistical inference of the
intractable estimations \cite{bishop2006pattern}. The core idea of
variational inference is to compute an approximate posterior PDF,
denoted as the surrogate PDF, $q(\boldsymbol{\mu})$, which can approximate
the true posterior PDF \eqref{eq:posteriorPDF} well. 

\subsection{Variational Inference}

According to the approximation inference theory \cite{bishop2006pattern},
the logarithm of the model evidence $p(\mathbf{y})$ can be decomposed
as
\begin{equation}
\ln p(\mathbf{y})=\mathrm{\mathcal{D}_{\mathrm{KL}}}(q\|p)+\mathcal{L}(q),\label{eq:logModelEvidence}
\end{equation}
where $\mathrm{\mathcal{D}_{\mathrm{KL}}}(q\|p)$ represents the Kullback-Leibler
(KL) divergence between the surrogate PDF $q(\boldsymbol{\mu})$ and
the true posterior PDF $p(\boldsymbol{\mu}\mid\mathbf{y})$. It is
defined as
\begin{equation}
\mathrm{\mathcal{D}_{\mathrm{KL}}}(q\|p)=-\int q(\boldsymbol{\mu})\ln\left\{ \frac{p(\boldsymbol{\mu}\mid\mathbf{y})}{q(\boldsymbol{\mu})}\right\} \mathrm{d}\boldsymbol{\mu}.\label{eq:KLdivergence}
\end{equation}
The function $\mathcal{L}(q)$ is defined as
\begin{equation}
\mathcal{L}(q)=\int q(\mathbf{\boldsymbol{\mu}})\ln\left\{ \frac{p(\mathbf{y},\mathbf{\boldsymbol{\mu}})}{q(\boldsymbol{\mu})}\right\} \mathrm{d}\boldsymbol{\mu}=\mathrm{E}_{q(\boldsymbol{\mu})}\left\{ \ln\frac{p(\mathbf{y},\boldsymbol{\mu})}{q(\boldsymbol{\mu})}\right\} .\label{eq:funtionalL}
\end{equation}
If we allow any possible choice for the surrogate PDF $q(\boldsymbol{\mu})$,
the KL divergence will satisfy $\mathrm{\mathcal{D}_{\mathrm{KL}}}(q\|p)\geqslant0$,
with equality if and only if the surrogate PDF equals the posterior
PDF $q(\boldsymbol{\mu})=p(\boldsymbol{\mu}\mid\mathbf{y})$. Consequently,
we obtain $\mathcal{L}(q)\leqslant\ln p(\mathbf{y})$ from \eqref{eq:logModelEvidence},
which means that $\mathcal{L}(q)$ is a lower bound on $\ln p(\mathbf{y})$.
This lower bound is known as the \textcolor{black}{Evidence Lower BOund
(ELBO)}. Although minimizing the KL divergence $\mathrm{\mathcal{D}_{\mathrm{KL}}}(q\|p)$
is equivalent to maximizing $\mathcal{L}(q)$, the actual posterior
PDF $p(\boldsymbol{\mu}\mid\mathbf{y})$ is generally intractable.

Therefore, we restrict our search to a tractable family of candidate
PDFs and seek the member within this family that maximizes the objective
$\mathcal{L}(q)$. The guiding principle is to ensure analytical tractability
while retaining sufficient flexibility to closely approximate the
true posterior PDF. To this end, we model the surrogate PDF $q(\boldsymbol{\mu})$
as a multivariate (real) Gaussian distribution, i.e.,
\begin{equation}
q(\boldsymbol{\mu})=f_{\mathrm{N}}\left(\boldsymbol{\mu};\hat{\boldsymbol{\mu}},\mathbf{\hat{V}}_{\boldsymbol{\mu}}\right),\label{eq:multGaussian}
\end{equation}
where \textcolor{black}{$\boldsymbol{\mu}\sim\mathcal{N}(\hat{\boldsymbol{\mu}},\mathbf{\hat{V}}_{\boldsymbol{\mu}})$},
$\hat{\boldsymbol{\mu}}=(\hat{\theta},\hat{r},\hat{g},\hat{\phi})^{T}$
is the set of unobserved variable estimates, and the confidence level
of the estimates $\hat{\mathbf{V}}_{\boldsymbol{\mu}}$ is given by
\[
\hat{\mathbf{V}}_{\boldsymbol{\mu}}=\left[\begin{array}{cccc}
\hat{v}_{\theta\theta} & \hat{v}_{\theta r} & \hat{v}_{\theta g} & \hat{v}_{\theta\phi}\\
\hat{v}_{\theta r} & \hat{v}_{rr} & \hat{v}_{rg} & \hat{v}_{r\phi}\\
\hat{v}_{\theta g} & \hat{v}_{rg} & \hat{v}_{gg} & \hat{v}_{g\phi}\\
\hat{v}_{\theta\phi} & \hat{v}_{r\phi} & \hat{v}_{g\phi} & \hat{v}_{\phi\phi}
\end{array}\right].
\]
The properties of the Gaussian distribution imply that $\mathrm{E}_{q(\boldsymbol{\mu})}\left\{ \ln q(\boldsymbol{\mu})\right\} $ is
a constant; consequently, the ELBO can be expressed as
\begin{equation}
\mathcal{L}(q)=\mathrm{E}_{q(\boldsymbol{\mu})}\left\{ \ln p(\mathbf{y},\boldsymbol{\mu})\right\} +\mathrm{const}.\label{eq:LwithGaussian}
\end{equation}

In practice, there is little or no knowledge about the channel information,
such as location and fading. \textcolor{black}{Hence, a flat (i.e.,
marginally uniform, with independence among involved parameters) prior
is used in this non-informative condition with minimal influence on
outcomes of posterior inference \cite{gelman1995bayesian}. This is
tantamount to assuming} $p(\mathbf{\boldsymbol{\mu}})=p(\theta)p(r)p(g)p(\phi)$,
where $p(\theta)=1/\text{\ensuremath{\pi}},\theta\in(0,\pi)$, $p(r)=1/r_{R},r\in(0,r_{R})$,
$p(g)=1/\sqrt{p_{t}},g\in(0,\text{\ensuremath{\sqrt{p_{t}}}})$ and
$p(\phi)=1/2\pi,\phi\in(0,2\pi)$. \textcolor{black}{Although the propagation
parameters are physically bounded, the massive number of antennas
in UM-MIMO systems causes the posterior distribution to become highly
peaked. Consequently, the boundary effects are negligible, making
the unconstrained Gaussian surrogate a standard and highly accurate
approximation, mathematically akin to the Laplace approximation \cite{bishop2006pattern}.
}Under this assumption, the prior PDF is not involved in the joint
PDF \eqref{eq:jointPDF}, i.e.,
\begin{equation}
p(\mathbf{y},\mathbf{\boldsymbol{\mu}})\wasypropto\prod_{m=1}^{M}f_{\mathrm{CN}}\left(y_{m};ge^{j\phi}e^{jk_{c}\left(r_{m}-r\right)},\sigma^{2}\right).\label{eq:factorJointPDF}
\end{equation}
Plugging the factor of joint PDF \eqref{eq:factorJointPDF} into
the ELBO \eqref{eq:LwithGaussian} yields
\begin{equation}
\begin{aligned}\mathcal{L}(q)= & \mathrm{E}_{q(\boldsymbol{\mu})}\left\{ \sum_{m=1}^{M}\ln f_{\mathrm{CN}}\left(y_{m};ge^{j\phi}e^{jk_{c}\left(r_{m}-r\right)},\sigma^{2}\right)\right\} \\
 & +\mathrm{const}.
\end{aligned}
\label{eq:LwithJointPDF}
\end{equation}
For a \textcolor{black}{complex Gaussian variable $z\sim\mathcal{CN}(\mu_{z},\sigma_{z}^{2})$},
\textcolor{black}{its PDF is given by} \textcolor{black}{$f_{\mathrm{CN}}(z;\mu_{z},\sigma_{z}^{2})=\frac{1}{\pi\sigma_{z}^{2}}e^{-\frac{|z-\mu_{z}|^{2}}{\sigma_{z}^{2}}}$}.
Based on \eqref{eq:LwithJointPDF}, maximizing the ELBO is equivalent
to maximizing \textcolor{black}{
\begin{equation}
f(\mathbf{\boldsymbol{\mu}})=\frac{1}{\sigma^{2}}\left(\sum_{m=1}^{M}2|y_{m}|g\cos(\psi)-Mg^{2}\right),\label{eq:optFunc}
\end{equation}
}where
\begin{equation}
\psi=k_{c}\left(r_{m}-r\right)+\phi-\angle y_{m}.
\end{equation}

For the function $f(\mathbf{\boldsymbol{\mu}})$, \textcolor{black}{the
components of the gradient vector with respect to} $(\theta,r,g,\phi)$
are\begin{subequations} \label{firstDerivative}\textcolor{black}{
\begin{equation}
\frac{\partial f(\mathbf{\boldsymbol{\mu}})}{\partial\theta}=\sum_{m=1}^{M}\frac{2k_{c}\delta_{m}d\left|y_{m}\right|gr}{\sigma^{2}r_{m}}\sin\psi,
\end{equation}
\begin{equation}
\frac{\partial f(\mathbf{\boldsymbol{\mu}})}{\partial r}=\sum_{m=1}^{M}-\frac{2k_{c}\left|y_{m}\right|g\sin\psi}{\sigma^{2}}\left(\frac{r-\delta_{m}d\cos\theta}{r_{m}}-1\right),
\end{equation}
\begin{equation}
\frac{\partial f(\mathbf{\boldsymbol{\mu}})}{\partial g}=\frac{1}{\sigma^{2}}\left(\sum_{m=1}^{M}2\left|y_{m}\right|\cos\psi-2Mg\right),\label{eq:1dg}
\end{equation}
\begin{equation}
\frac{\partial f(\mathbf{\boldsymbol{\mu}})}{\partial\phi}=\sum_{m=1}^{M}-\frac{2\left|y_{m}\right|g\sin\psi}{\sigma^{2}}.\label{eq:1dphi}
\end{equation}
}\end{subequations} Our goal is to maximize the function $f(\mathbf{\boldsymbol{\mu}})$
by updating $\mathbf{\boldsymbol{\mu}}$, but updating the four parameters
in $\mathbf{\boldsymbol{\mu}}$ simultaneously is formidable. To reduce
the burden of the optimization, we set the first-order derivatives
\eqref{eq:1dg} and \eqref{eq:1dphi} equal to zero, and find that
the estimate of the complex channel gain that maximizes the function
$f(\mathbf{\boldsymbol{\mu}})$ is also the corresponding \textcolor{black}{Least
Squares (LS)} solution, given by
\begin{equation}
\hat{g}e^{j\hat{\phi}}=\frac{\mathbf{b}(\theta,r)^{H}\mathbf{y}}{\left\Vert \mathbf{b}(\theta,r)\right\Vert ^{2}}.\label{eq:LSgain}
\end{equation}
\textcolor{black}{By substituting the LS solution of the channel gain
$\hat{g}e^{j\hat{\phi}}$ from \eqref{eq:LSgain} back into the measurement
model, we can evaluate the energy of the residual signal, given by
$||y-\hat{g}e^{j\hat{\phi}}b(\theta,r)||^{2}=||y||^{2}-\frac{|b(\theta,r)^{H}y|^{2}}{||b(\theta,r)||^{2}}$.
It is evident that minimizing this residual energy is mathematically
equivalent to maximizing the term $\frac{|b(\theta,r)^{H}y|^{2}}{||b(\theta,r)||^{2}}$.
Consequently, by} replacing the channel gain $\hat{g}e^{j\hat{\phi}}$
with a fixed value \eqref{eq:LSgain} in the function $f(\mathbf{\boldsymbol{\mu}})$
\eqref{eq:optFunc}, we can derive that the estimate of the propagation
parameters $(\hat{\theta},\hat{r})$ are obtained as the solution
to the subsequent optimization problem,
\begin{equation}
(\hat{\theta},\hat{r})=\arg\max_{\theta,r}G_{\mathbf{y}}(\theta,r),\label{eq:GLRT}
\end{equation}
where the cost function $G_{\mathbf{y}}(\theta,r)$ is given by
\begin{equation}
G_{\mathbf{y}}(\theta,r)=\frac{\left|\mathbf{b}\left(\theta,r\right)^{H}\mathbf{y}\right|^{2}}{\left\Vert \mathbf{b}(\theta,r)\right\Vert ^{2}}.\label{eq:costFunc}
\end{equation}
Consequently, our task reduces to designing an efficient optimization
algorithm to jointly refine the propagation estimates $(\theta,r)$
while iteratively updating the estimated channel gain $\hat{g}e^{j\hat{\phi}}$
so as to maximize the original objective function $f(\mathbf{\boldsymbol{\mu}})$. 

\textcolor{black}{Because the highly non-linear nature of the near-field
spatial steering vectors makes standard conjugate variational updates
intractable, we resort to a Laplace approximation within our variational
framework. Specifically, we utilize deterministic Newton optimization
to find the mode of the local objective function, and subsequently
approximate the local posterior as a Gaussian distribution, where
the covariance is obtained from the inverse Hessian matrix
evaluated at this mode.}

\subsection{Newton Gradient Descent\label{subsec:Newton-Gradient-Descent}}

The Newton gradient descent method is utilized in our algorithm to
maximize the function $f(\mathbf{\boldsymbol{\mu}})$ by updating
the estimate $\hat{\boldsymbol{\mu}}$ and its confidence level $\mathbf{\hat{V}}_{\boldsymbol{\mu}}$.
Since the optimal solution for the gain parameters $(\hat{g},\hat{\phi})$
is already given by \eqref{eq:LSgain}, the Newton optimization step
for estimate update is only applied to the position parameters $(\hat{\theta},\hat{r})$.
Specifically, we update the estimates as\textcolor{black}{
\begin{equation}
(\hat{\theta}',\hat{r}')^{T}\triangleq(\hat{\theta},\hat{r})^{T}-[\nabla^{2}f(\hat{\theta},\hat{r})]^{-1}\nabla f(\hat{\theta},\hat{r}),\label{eq:estimateUpdate}
\end{equation}
}where\textcolor{black}{
\begin{equation}
\nabla f(\theta,r)=\begin{bmatrix}\frac{\partial f(\mu)}{\partial\theta}\\
\frac{\partial f(\mu)}{\partial r}
\end{bmatrix},\label{eq:firstDerivativeVector}
\end{equation}
\begin{equation}
\nabla^{2}f(\theta,r)=\begin{bmatrix}\frac{\partial^{2}f(\mu)}{\partial\theta^{2}} & \frac{\partial^{2}f(\mu)}{\partial\theta\partial r}\\
\frac{\partial^{2}f(\mu)}{\partial\theta\partial r} & \frac{\partial^{2}f(\mu)}{\partial r^{2}}
\end{bmatrix}.\label{eq:secondDerivativeMatrix}
\end{equation}
}are the \textcolor{black}{gradient vector} and the \textcolor{black}{Hessian
matrix} of the function $f(\mathbf{\boldsymbol{\mu}})$, respectively.
The expressions for the \textcolor{black}{components of the gradient
vector} are defined in \eqref{firstDerivative}, and the \textcolor{black}{elements
of the Hessian matrix} are shown in \eqref{secondDerivative}. 
\begin{figure*}[tbh]
 \begin{subequations} \label{secondDerivative}

\textcolor{black}{
\begin{equation}
\frac{\partial^{2}f(\mathbf{\boldsymbol{\mu}})}{\partial\theta^{2}}=\sum_{m=1}^{M}\frac{2k_{c}\delta_{m}^{2}d^{2}\left|y_{m}\right|gr^{2}}{\sigma^{2}r_{m}^{2}}\left(\frac{\sin\psi}{r_{m}}-k_{c}\cos\psi\right),
\end{equation}
\begin{equation}
\frac{\partial^{2}f(\mathbf{\boldsymbol{\mu}})}{\partial\theta\partial r}=\sum_{m=1}^{M}\frac{2k_{c}\delta_{m}d\left|y_{m}\right|g}{\sigma^{2}r_{m}}\left(k_{c}\left(\frac{r-\delta_{m}d\cos\theta}{r_{m}}-1\right)\cos\psi-\left(\frac{r^{2}-\delta_{m}dr\cos\theta}{r_{m}^{2}}-1\right)\sin\psi\right),
\end{equation}
\begin{equation}
\frac{\partial^{2}f(\mathbf{\boldsymbol{\mu}})}{\partial\theta\partial g}=\sum_{m=1}^{M}\frac{2k_{c}\delta_{m}d\left|y_{m}\right|r}{\sigma^{2}r_{m}}\sin\psi,\ \frac{\partial^{2}f(\mathbf{\boldsymbol{\mu}})}{\partial\theta\partial\phi}=\sum_{m=1}^{M}\frac{2k_{c}\delta_{m}d\left|y_{m}\right|gr}{\sigma^{2}r_{m}}\cos\psi.
\end{equation}
\begin{equation}
\frac{\partial^{2}f(\mathbf{\boldsymbol{\mu}})}{\partial r^{2}}=\sum_{m=1}^{M}\frac{2k_{c}\left|y_{m}\right|g}{\sigma^{2}}\left(\left(\frac{\left(r-\delta_{m}d\cos\theta\right)^{2}}{r_{m}^{3}}-\frac{1}{r_{m}}\right)\sin\psi-k_{c}\left(\frac{r-\delta_{m}d\cos\theta}{r_{m}}\right)^{2}\cos\psi\right)
\end{equation}
\begin{equation}
\frac{\partial^{2}f(\mathbf{\boldsymbol{\mu}})}{\partial r\partial g}=\sum_{m=1}^{M}-\frac{2k_{c}\left|y_{m}\right|}{\sigma^{2}}\left(\frac{r-\delta_{m}d\cos\theta}{r_{m}}-1\right)\sin\psi,\ \frac{\partial^{2}f(\mathbf{\boldsymbol{\mu}})}{\partial r\partial\phi}=\sum_{m=1}^{M}-\frac{2k_{c}\left|y_{m}\right|g}{\sigma^{2}}\left(\frac{r-\delta_{m}d\cos\theta}{r_{m}}-1\right)\cos\psi,
\end{equation}
\begin{equation}
\frac{\partial^{2}f(\mathbf{\boldsymbol{\mu}})}{\partial g^{2}}=-\frac{2M}{\sigma^{2}},\ \frac{\partial^{2}f(\mathbf{\boldsymbol{\mu}})}{\partial g\partial\phi}=\sum_{m=1}^{M}-\frac{2\left|y_{m}\right|\sin\psi}{\sigma^{2}},\ \frac{\partial^{2}f(\mathbf{\boldsymbol{\mu}})}{\partial\phi^{2}}=\sum_{m=1}^{M}-\frac{2\left|y_{m}\right|g\cos\psi}{\sigma^{2}}.
\end{equation}
}

\hrule\end{subequations}
\end{figure*}
 
\begin{algorithm}[tbh]
\caption{Near-Field Newton Gradient Descent \label{alg:NewtonUpdate}}

\textbf{Input:} Measurement $\mathbf{y}$, Original estimate $\hat{\boldsymbol{\mu}}=(\hat{\theta},\hat{r},\hat{g},\hat{\phi})$;

\begin{algorithmic}[1]

\Statex // Locally Concave Condition

\If {\textcolor{black}{$\nabla^{2}f(\hat{\theta},\hat{r})$ }is negative
definite}

\State Perform the Newton optimization step update in \eqref{eq:estimateUpdate}:
\textcolor{black}{
\[
(\hat{\theta}',\hat{r}')^{T}\leftarrow(\hat{\theta},\hat{r})^{T}-[\nabla^{2}f(\hat{\theta},\hat{r})]^{-1}\nabla f(\hat{\theta},\hat{r});
\]
}

\Statex // Near-Field Restriction Condition

\State $\hat{r}'\leftarrow\min(\hat{r}',r_{R})$;

\Statex // Residual Energy Decrease Condition

\If {$G_{\mathbf{y}}(\hat{\theta},\hat{r})>G_{\mathbf{y}}(\hat{\theta}',\hat{r}')$}

\State $(\hat{\theta}',\hat{r}')\leftarrow(\hat{\theta},\hat{r})$;

\EndIf

\Else

\State $(\hat{\theta}',\hat{r}')\leftarrow(\hat{\theta},\hat{r})$;

\EndIf

\State Perform the corresponding LS solution in \eqref{eq:LSgain}:\textcolor{black}{
\[
\hat{g}'e^{j\hat{\phi}'}\leftarrow\mathbf{b}(\hat{\theta}',\hat{r}')^{H}\mathbf{y}/||\mathbf{b}(\hat{\theta}',\hat{r}')||^{2};
\]
}

\State Perform the confidence level update in \eqref{eq:varianceUpdate}:
\[
\mathbf{\hat{V}}_{\boldsymbol{\mu}}'\gets\mathtt{-\mathbf{H}'^{-1}};
\]

\end{algorithmic}

\textbf{Output:} Updated estimate $\hat{\boldsymbol{\mu}}'=(\hat{\theta}',\hat{r}',\hat{g}',\hat{\phi}')$,
Updated confidence level $\mathbf{\hat{V}}_{\boldsymbol{\mu}}'$.
\end{algorithm}

The near-field Newton gradient descent procedure is illustrated in
\textbf{Algorithm \ref{alg:NewtonUpdate}}. In addition to the general
Newton optimization step described in Step 2, we have incorporated
three near-field conditions to ensure that the optimization is conducted
in the desired manner:
\begin{enumerate}
\item \textit{Locally Concave}: Since the objective is to maximize the function
$f(\mathbf{\boldsymbol{\mu}})$, Step 1 specifies that the Newton
optimization step \eqref{eq:estimateUpdate} is performed only when
the function is locally concave at $(\hat{\theta},\hat{r})$, i.e.,
\textcolor{black}{$\nabla^{2}f(\hat{\theta},\hat{r})$} is negative
definite.
\item \textit{Near-Field Restriction}: As the function $f(\mathbf{\boldsymbol{\mu}})$
is derived from \textcolor{black}{the} near-field model, it is essential to ensure that the
estimated distance $\hat{r}$ does not exceed the Rayleigh distance
$r_{R}$. Step 3 introduces an additional rule to restrict the distance
after the Newton optimization step.
\item \textit{Residual Energy Decrease}: To guarantee the convergence of
our algorithm, it is necessary to observe a decrease in the overall
residual energy. Hence, Steps 4-6 specify that we only accept the
update if the residual energy is non-increasing, i.e., the cost function
in \eqref{eq:costFunc} is not less than the original. Otherwise,
the current estimate is retained. 
\end{enumerate}
\par In Step 10, after the Newton optimization step, the complex
channel coefficient is updated using the LS solution in \eqref{eq:LSgain}.
Furthermore, since the surrogate PDF $q(\boldsymbol{\mu})$ is constrained
to follow a multivariate Gaussian distribution \eqref{eq:multGaussian},
the update rule for the confidence level $\hat{\mathbf{V}}_{\boldsymbol{\mu}}$
is given by
\begin{equation}
\hat{\mathbf{V}}{}_{\boldsymbol{\mu}}=\mathtt{-\mathbf{\mathbf{H}}^{-1}},\label{eq:varianceUpdate}
\end{equation}
where \textcolor{black}{$\mathbf{H}=\nabla^{2}f(\mathbf{\boldsymbol{\mu}})$
is the aforementioned} Hessian matrix. Step 11 involves updating the
confidence level with the Hessian matrix $\mathbf{H}'$ of the function
$f(\hat{\theta}',\hat{r}',\hat{g}',\hat{\phi}')$. \textcolor{black}{According
to }\textbf{\textcolor{black}{Lemma 1}}\textcolor{black}{, this Hessian
matrix is guaranteed to be strictly negative definite upon convergence,
which rigorously ensures the mathematical validity of the confidence
level update.}
\begin{lem}
\textcolor{black}{In the proposed near-field channel estimation model
based on the variable projection method, if $\nabla^{2}f(\hat{\theta},\hat{r})$
is strictly negative definite at a stationary point, then the full
Hessian matrix $\mathbf{H}$ evaluated at the reconstructed parameter
coordinates is strictly negative definite. This property mathematically
guarantees that the extracted covariance matrix $\hat{\mathbf{V}}_{\boldsymbol{\mu}}=-\mathbf{H}^{-1}$
is strictly positive definite.}
\end{lem}
\begin{IEEEproof}
\textcolor{black}{See Appendix. }
\end{IEEEproof}
In the preceding subsections of this section, we derive the Bayesian
formulation, variational inference, and Newton gradient descent algorithm
for the single-path case, i.e., $L=1$. However, these methods are
also effective in the multiple-path model. The general VNNCE algorithm,
which is a direct extension of the single-path optimization to the
multiple-path case, will be discussed in the next subsection. 

\subsection{VNNCE Algorithm for Multiple Paths}

In the multiple-path scenario, we denote the unknown channel parameters
of the $l$th path as $\mathbf{\boldsymbol{\mu}}_{l}=(\theta_{l},r_{l},g_{l},\phi_{l})^{T}$,
and a set of unknown channel parameters for different paths as $\mathbf{\boldsymbol{\mu}}_{1:k}=\{\theta_{l},r_{l},g_{l},\phi_{l}:l=1,2,\ldots,k\}$.
Assuming that the first $k$ paths have been estimated, the residual
measurement corresponding to this set of unknown channel parameters
is
\begin{equation}
\mathbf{y}_{\mathrm{r}}(\mathbf{\boldsymbol{\mu}}_{1:k})=\mathbf{y}-\sum_{l=1}^{k}g_{l}e^{j\phi_{l}}\mathbf{b}\left(\theta_{l},r_{l}\right).\label{eq:RNSM}
\end{equation}
As one of the off-grid estimation methods, we divide the estimation
procedure into two stages: the on-grid detection stage and the off-grid
refinement stage.

The detection stage aims to obtain a coarse estimate of the unknown
channel parameters by restricting them to a discrete set, which serves
as an initial guess for the refinement stage. To this end, we employ
the \textcolor{black}{Orthogonal Matching Pursuit (OMP)} algorithm
\cite{mallat1993matching,bajwa2010compressed}, a classical
sparse recovery method, to obtain a coarse estimate of the propagation
parameters $(\theta,r)$ by constraining them to a \textcolor{black}{Two-Dimensional
(2D)} finite discrete codebook $\Omega$. In practice, the quality
of the coarse estimate provided by the detection stage critically
influences the convergence and accuracy of the refinement stage; consequently,
a well-designed codebook is of paramount importance. The structure
of the codebook is defined as $\Omega=\{(\theta_{n_{a}},r_{n_{d}}):n_{a}=1,2,\ldots,N_{a},\ n_{d}=1,2,\ldots,N_{d}\}$
and will be elaborated in detail in Sec. \ref{sec:Codebook-Design}.
For the $l$th path corresponding to the residual measurement, the
coarse estimate $(\hat{\theta}_{l},\hat{r}_{l})$ obtained by the
OMP algorithm is given by
\begin{equation}
(\hat{\theta}_{l},\hat{r}_{l})=\arg\max_{(\theta,r)\in\Omega}\left|\mathbf{b}^{H}(\theta,r)\mathbf{y}_{\mathrm{r}}(\mathbf{\boldsymbol{\mu}}_{1:l-1})\right|^{2},\label{eq:OMPcodition}
\end{equation}
which is equivalent to maximizing the cost function $G_{\mathbf{y}_{r}}(\theta,r)$
in \eqref{eq:costFunc}. Subsequently, the corresponding complex channel
coefficient can be obtained as $\hat{g}_{l}e^{j\hat{\phi}_{l}}=\mathbf{b}^{H}(\theta,r)\mathbf{y}_{\mathrm{r}}(\mathbf{\boldsymbol{\mu}}_{1:l-1})/M$.

To address the estimation error introduced by the discrete estimates
from the detection stage, the refinement stage is proposed to extend
the Newton gradient descent from a single path to multiple paths.
The refinement stage maximizes the ELBO over the continuum rather
than discrete points, using the Newton gradient descent algorithm
described in \textbf{Algorithm \ref{alg:NewtonUpdate}}. Given the
fixed paths characterized by $\{(\hat{\boldsymbol{\mu}}_{l},\mathbf{\hat{V}}_{l}),l=1,2,\ldots,i-1,i+1,i+2,\ldots,k\}$,
the joint PDF of $i$th path's unknown channel parameters is given
by
\begin{align}
p(\mathbf{y},\mathbf{\boldsymbol{\mu}}_{i}) & \wasypropto\prod_{m=1}^{M}f_{\mathrm{CN}}\left(y_{m};\Xi+g_{i}e^{j\phi_{i}}e^{jk_{c}\left(r_{i,m}-r_{i}\right)},\sigma^{2}\right)\nonumber \\
 & \wasypropto\prod_{m=1}^{M}f_{\mathrm{CN}}\left(y_{m}-\Xi;g_{i}e^{j\phi_{i}}e^{jk_{c}\left(r_{i,m}-r_{i}\right)},\sigma^{2}\right),\label{eq:jointPDFmult}
\end{align}
where $\Xi$ is the sum of all fixed channel \textcolor{black}{contributions}, i.e.,
\begin{equation}
\Xi=\sum_{l=1,l\neq i}^{k}g_{l}e^{j\phi_{l}}e^{jk_{c}\left(r_{l,m}-r_{l}\right)}.\label{eq:sumFixedChannel}
\end{equation}
The joint PDF \eqref{eq:jointPDFmult} shows that \textcolor{black}{$y_{m}-\Xi$},
which \textcolor{black}{follows a complex Gaussian distribution}, represents
the $m$th element of the vector $\mathbf{y}_{\mathrm{r}}(\mathbf{\boldsymbol{\mu}}_{1:i-1,i+1:k})$.
Therefore, the Newton gradient descent in the multiple-path case can
be performed similarly to the single-path case, as discussed in Sec.
\ref{subsec:Newton-Gradient-Descent}. 
\begin{algorithm}[t]
\caption{Proposed VNNCE Algorithm \label{alg:NVINCE}}

\textbf{Input:} Measurement $\mathbf{y}$, Numbers of rounds $R_{s}$
and $R_{c}$, Number of channel paths $L'$;

\textbf{Initialization:} $l\gets0$;

\begin{algorithmic}[1]

\While {$l<L'$}

\State $l\gets l+1;$

\Statex // Coarse Estimate 

\State Perform the OMP algorithm in \eqref{eq:OMPcodition}: 
\[
(\hat{\theta}_{l},\hat{r}_{l})\text{\ensuremath{\gets}}\arg\max_{(\theta,r)\in\Omega}G_{\mathbf{y}_{\mathrm{r}}(\mathbf{\boldsymbol{\mu}}_{1:l-1})}(\theta,r);
\]

\State Perform the corresponding \textcolor{black}{LS} estimate in \eqref{eq:LSgain}:\textcolor{black}{
\[
\hat{g}_{l}e^{j\hat{\phi}_{l}}\leftarrow\mathbf{b}(\hat{\theta}_{l},\hat{r}'_{l})^{H}\mathbf{y}_{\mathrm{r}}(\mathbf{\boldsymbol{\mu}}_{1:l-1})/||\mathbf{b}(\hat{\theta}_{l},\hat{r}'_{l})||^{2};
\]
}

\Statex // Single Refinement

\State Fixed $\mathbf{\boldsymbol{\hat{\mu}}}_{1:l-1}$,i.e.,we treat
$\mathbf{y}_{\mathrm{r}}(\mathbf{\boldsymbol{\hat{\mu}}}_{1:l-1})$
as the measurement, refine $R_{s}$ rounds of Near-field Newton gradient
descent in \textbf{\textcolor{black}{Algorithm}} \ref{alg:NewtonUpdate}
on $\hat{\boldsymbol{\mu}}_{l}$ to obtain $\hat{\boldsymbol{\mu}}_{l}'$
and $\hat{\mathbf{V}}_{\boldsymbol{\mu},l}'$:

\State $\mathbf{\boldsymbol{\hat{\mu}}}_{1:l}\gets\mathbf{\boldsymbol{\hat{\mu}}}_{1:l-1}\cup\{\hat{\boldsymbol{\mu}}_{l}'\},\ \hat{\boldsymbol{\mu}}_{l}\gets\hat{\boldsymbol{\mu}}_{l}',\ \hat{\mathbf{V}}_{\boldsymbol{\mu},l}\gets\hat{\mathbf{V}}_{\boldsymbol{\mu},l}';$

\Statex // Cyclic Refinement

\For{$i=1:R_{c}$}

\For{each $\hat{\boldsymbol{\mu}}_{k}$ in $\mathbf{\boldsymbol{\hat{\mu}}}_{1:l}$}

\State Fixed $\mathbf{\boldsymbol{\hat{\mu}}}_{1:k-1,k+1:l}$, i.e.,
we treat $\mathbf{y}_{\mathrm{r}}(\mathbf{\boldsymbol{\mu}}_{1:k-1,k+1:l})$
as the measurement, refine $R_{s}$ rounds of Near-field Newton gradient
descent in \textbf{\textcolor{black}{Algorithm}} \ref{alg:NewtonUpdate}
on $\hat{\boldsymbol{\mu}}_{k}$ to obtain $\hat{\boldsymbol{\mu}}_{k}'$
and $\hat{\mathbf{V}}_{\boldsymbol{\mu},k}'$:

\State $\mathbf{\boldsymbol{\hat{\mu}}}_{1:l}\gets\mathbf{\boldsymbol{\hat{\mu}}}_{1:k-1}\cup\{\hat{\boldsymbol{\mu}}_{k}'\}\cup\mathbf{\boldsymbol{\hat{\mu}}}_{k+1:l},\ \hat{\boldsymbol{\mu}}_{k}\gets\hat{\boldsymbol{\mu}}_{k}',\ \hat{\mathbf{V}}_{\boldsymbol{\mu},k}\gets\hat{\mathbf{V}}_{\boldsymbol{\mu},k}';$

\EndFor

\EndFor

\EndWhile

\end{algorithmic}

\textbf{Output:} Estimated result of channel parameters and their
confidence levels $\{\mathbf{\boldsymbol{\hat{\mu}}}_{l},\hat{\mathbf{V}}_{\boldsymbol{\mu},l}:l=1,2,\ldots,L'\}$.
\end{algorithm}

The VNNCE algorithm for the multi-path case is summarized in \textbf{Algorithm
\ref{alg:NVINCE}}. The main components of the algorithm are discussed
as follows:
\begin{enumerate}
\item \textit{Coarse Estimate}: As mentioned in the detection stage, Steps
3 and 4 provide the coarse estimates as the initial guess for optimization.
\item \textit{Single Refinement}: In Step 5, $R_{s}$ rounds of near-field
Newton gradient descent are utilized to refine $\mathbf{\boldsymbol{\hat{\mu}}}_{l}$
and $\hat{\mathbf{V}}_{\boldsymbol{\mu},l}$ of the $l$th path.
\item \textit{Cyclic Refinement}: In Steps 7-12, $R_{c}$ rounds of \textit{Single
Refinement} are employed for further refinements. Unlike conventional
forward algorithms, this step provides feedback for optimization\textcolor{black}{.
Inspired by the NOMP framework in \cite{mamandipoor2016newtonized},
this cyclic refinement procedure is specifically designed to iteratively
reconstruct and subtract the interference of other paths, thereby
effectively decoupling the inter-path interference and improving estimation
accuracy.}
\end{enumerate}

\subsection{Complexity Analysis}

We analyze the computational complexity of VNNCE algorithm in \textbf{Algorithm
\ref{alg:NVINCE}}. In the \textit{Coarse Estimate} step, the cost
function in \eqref{eq:costFunc} is computed over the 2D discrete
codebook $\Omega$ with size $S$, and \textcolor{black}{the complexity of this step is} $\mathcal{O}(L^{\prime}SM)$. The \textit{Single Refinement}
step only requires $\mathcal{O}(L^{\prime}R_{s}M)$ operations for
the entire algorithm. In contrast, the \textit{Cyclic Refinement}
step involves refining all paths that have been estimated so far,
and has overall complexity $\mathcal{O}(L^{\prime2}R_{c}R_{s}M)$.
Empirical observations suggest that, unlike the NOMP algorithm in
\cite{mamandipoor2016newtonized}, which is a typical far-field off-grid
channel estimation algorithm where the Cyclic Refinement step dominates
the overall computational cost, the Coarse Estimate step dominates
the overall computational cost of the VNNCE algorithm. 
\begin{table}[t]
\caption{Average Execution Time Comparison \label{tab:CCC}}

\centering{}%
\begin{tabular}{|l|l|}
\hline 
\textbf{Algorithm} & \textbf{Computational Complexity $\mathcal{O}(\cdot)$}\tabularnewline
\hline 
VNNCE & $\mathcal{O}(L_{1}^{\prime}S_{1}M)+\mathcal{O}(L_{1}^{\prime2}R_{c}R_{s}M)$\tabularnewline
\hline 
P-SOMP \cite{cui2022channel} & $\mathcal{O}(L_{2}^{\prime}S_{2}M)$\tabularnewline
\hline 
P-SIGW \cite{cui2022channel} & $\mathcal{O}(L_{2}^{\prime}S_{2}M)+\mathcal{O}(N_{\textrm{iter}}(L_{2}^{'3}+L_{2}^{'2}M+L_{2}^{'}M^{2}))$\tabularnewline
\hline 
N-NOMP \cite{lu2022near} & $\mathcal{O}(L_{3}^{\prime}S_{3}M)+\mathcal{O}(L_{3}^{\prime2}R_{c}R_{s}M)$\tabularnewline
\hline 
\end{tabular}
\end{table}

The computational complexities of existing near-field channel estimation
algorithms are compared and summarized in Table \ref{tab:CCC}. In
this table, $S_{1}$, $S_{2}$ and $S_{3}$ denote the sizes of the
codebooks used in the respective algorithms, while $L_{1}$, $L_{2}$
and $L_{3}$ represent the number of channel paths that need to be
identified. Additionally, $N_{\textrm{iter}}$ denotes the number
of line search iterations in the P-SIGW algorithm. Upon examining
the descriptions of these algorithms, we can observe certain relationships
among the parameters in the same scenario. Specifically, we have $M<S_{1}\approx S_{2}\ll S_{3}$,
$L'_{1}=L'_{3}<L'_{2}\ll M$, $R_{c},R_{s}\ll M$ and $N_{\textrm{iter}}\leqslant10$.
Based on these observations, it is clear that the proposed VNNCE algorithm
enjoys the best computational complexity among off-grid channel estimation
algorithms. The P-SIGW algorithm incurs a significant complexity cost
during the line search optimization step, which dominates the overall
computational complexity of the algorithm. On the other hand, although
our algorithm exhibits similar complexity to the N-NOMP algorithm,
we utilize a significantly smaller-scale codebook and reduce the complexity
of the Coarse Estimate step, thus dominating the overall computational
cost of the algorithm. In the other words, the VNNCE algorithm demonstrates
a comparable computational complexity to the on-grid P-SOMP algorithm.

\subsection{Cram\'{e}r-Rao Lower Bound}

\textcolor{black}{The} CRLB is a fundamental performance analysis tool that provides a theoretical
lower bound on the variance of any unbiased estimator, thereby serving
as a benchmark to evaluate the estimation performance of the proposed
algorithm. In this case, we consider parameter estimation in an \textcolor{black}{Additive
White Gaussian Noise (AWGN)} scenario, where the noisy observation
vector $\mathbf{y}$ can be described as
\begin{equation}
\mathbf{y}=\mathbf{s}(\boldsymbol{\mu})+\mathbf{z},\label{eq:AWGNy}
\end{equation}
with \textcolor{black}{$\mathbf{z}\sim\mathcal{CN}\left(\mathbf{0}_{M},\sigma^{2}\mathbf{I}_{M}\right)$}.
\textcolor{black}{Based on our near-field channel model in \eqref{eq:NSM},
$\boldsymbol{\mu}=\{\theta_{l},r_{l},g_{l},\phi_{l}:l=1,2,\ldots,L\}\in\mathbb{R}^{4L}$
is the joint vector aggregating all unknown channel parameters simultaneously,
and $\mathbf{s}(\boldsymbol{\mu})=\mathbf{h}=\sum_{l=1}^{L}\mathbf{h}_{l}=\sum_{l=1}^{L}g_{l}e^{j\phi_{l}}\mathbf{b}\left(\theta_{l},r_{l}\right)\in\mathbb{C}^{M}$
represents the noise-free received signal.} We assume an unbiased
estimate of $\boldsymbol{\alpha}^{T}\boldsymbol{\mu}$ as $\boldsymbol{\alpha}^{T}\boldsymbol{\hat{\mu}}(\mathbf{y})$,
where \textcolor{black}{$\boldsymbol{\alpha}\in\mathbb{R}^{4L}$} is
the weight vector. The variance of the estimator, given by $\mathbb{E}_{\mathbf{y}\mid\boldsymbol{\mu}}\left(\boldsymbol{\alpha}^{T}\hat{\boldsymbol{\mu}}(\mathbf{y})-\boldsymbol{\alpha}^{T}\boldsymbol{\mu}\right)^{2}$,
is lower bounded by $\boldsymbol{\alpha}^{T}F^{-1}(\boldsymbol{\mu})\boldsymbol{\alpha}$.
The $(i,j)$th elment in FIM is obtained as
\begin{equation}
F_{i,j}(\boldsymbol{\mu})=\mathbb{E}_{\mathbf{y}\mid\boldsymbol{\mu}}\left\{ \frac{\partial\ln p(\mathbf{y}\mid\boldsymbol{\mu})}{\partial\mu_{i}}\frac{\partial\ln p(\mathbf{y}\mid\boldsymbol{\mu})}{\partial\mu_{j}}\right\} .\label{eq:FIM}
\end{equation}
For the AWGN scenario in \eqref{eq:AWGNy}, \eqref{eq:FIM} can be
simplified to
\begin{equation}
F_{i,j}(\boldsymbol{\mu})=\frac{2}{\sigma^{2}}\Re\left\{ \left(\frac{\partial\mathbf{s}(\boldsymbol{\mu})}{\partial\mu_{i}}\right)^{H}\frac{\partial\mathbf{s}(\boldsymbol{\mu})}{\partial\mu_{j}}\right\} .\label{eq:FIMs}
\end{equation}

In the parameter estimation problem in \eqref{eq:NSM}, $\boldsymbol{\mu}$
consists of $\{\theta_{l},r_{l},g_{l},\phi_{l}:l=1,2,\ldots,L\}$,
and $\mathbf{s}(\boldsymbol{\mu})=\mathbf{h}=\sum_{l=1}^{L}\mathbf{h}_{l}=\sum_{l=1}^{L}g_{l}e^{j\phi_{l}}\mathbf{b}\left(\theta_{l},r_{l}\right)$.
For each $l$th single path, the $m$th element of \textcolor{black}{the
gradient vector} are\begin{subequations}
\begin{equation}
v_{\theta_{l}}^{m}=g_{l}e^{j\phi_{l}}e^{jk_{c}\left(r_{l,m}-r_{l}\right)}\times\frac{jk_{c}\delta_{m}dr_{l}\sin\theta_{l}}{r_{l,m}},
\end{equation}
\begin{equation}
v_{r_{l}}^{m}=g_{l}e^{j\phi_{l}}e^{jk_{c}\left(r_{l,m}-r_{l}\right)}\times\frac{jk_{c}(r_{l}-\delta_{m}d\cos\theta_{l}-r_{l,m})}{r_{l,m}},
\end{equation}
\begin{equation}
v_{g_{l}}^{m}=e^{j\phi_{l}}e^{jk_{c}\left(r_{l,m}-r_{l}\right)},
\end{equation}
\begin{equation}
v_{\phi_{l}}^{m}=jg_{l}e^{j\phi_{l}}e^{jk_{c}\left(r_{l,m}-r_{l}\right)}.
\end{equation}
\end{subequations} $F(\boldsymbol{\mu})$ is formed by \eqref{eq:FIMs},
and the CRLB of $\{\theta_{l},r_{l},g_{l},\phi_{l}:l=1,2,\ldots,L\}$
are given by the diagonal elements of $F^{-1}(\boldsymbol{\mu})$. 

\section{Codebook Design \label{sec:Codebook-Design}}

In this section, we derive the requirements for the near-field codebook
$\Omega$ design at the detection stage\textcolor{black}{. Unlike conventional
far-field codebooks, our near-field codebook is constructed using
steering vectors $\mathbf{b}(\theta,r)$ sampled across both angle
and distance dimensions to accurately capture the curvature of spherical
wavefronts. Our primary objective is to generate a highly reliable
initial guess for the VNNCE algorithm, facilitating the subsequent
Newton gradient descent to converge toward the true maximum of the
cost function in \eqref{eq:costFunc}.}

\textcolor{black}{Strictly speaking, proving the absolute convergence
of a joint 2D non-convex optimization problem requires multi-variable
bounds, which are analytically intractable for the highly non-linear
near-field steering vectors. To make the codebook design
tractable, we adopt a practical decoupled approach. By sequentially
fixing one variable to project the 2D landscape into independent One-Dimensional
(1D) slices, we utilize the 1D Newton convergence theory in }\textit{\textcolor{black}{Claim
1}}\textcolor{black}{{} as a practical guideline
to determine the grid spacing for each dimension independently. Because
this decoupled 1D heuristic cannot strictly guarantee joint 2D local
concavity at every initial grid point, we explicitly enforce the }\textit{\textcolor{black}{Locally
Concave}}\textcolor{black}{{} condition check in }\textbf{\textcolor{black}{Algorithm}}\textcolor{black}{{}
\ref{alg:NewtonUpdate}. If this condition is violated during the
refinement stage, the algorithm promptly terminates the iteration
to prevent divergence or performance degradation. This early-stopping
safeguard improves the robustness of the system in practice.}
\begin{claim}
For any function $f(x)$, we assume the initial guess $x_{0}$ falls
within an interval $\mathcal{I}=[x_{t}-\epsilon_{0},x_{t}+\epsilon_{0}]$
around the true solution $x_{t}$ where $\epsilon_{0}$ is a constant
and $f^{\prime}(x_{t})=0$. As mentioned in \cite{ascher2011first},
the Newton gradient descent method guarantees local convergence to
the true solution $x_{t}$, if the following three conditions are
satisfied:
\end{claim}
\begin{enumerate}
\item $\forall x\in\mathcal{I}$, $f^{\prime\prime}(x)\neq0$;
\item $\forall x\in\mathcal{I}$, $f^{\prime\prime\prime}(x)\ \textrm{is continuous}$;
\item Given $M\triangleq\frac{1}{2}(\sup_{x\in\mathcal{I}}|f^{\prime\prime\prime}(x)|)(\sup_{x\in\mathcal{I}}|1/f^{\prime\prime}(x)|)$,
$|\epsilon_{0}|<1/M$.
\end{enumerate}
For convenience, consider a single-path case as an example. We assume
the complex channel gain, angle and distance are $g_{t}e^{j\phi_{t}}$,
$\theta_{t}$ and $r_{t}$, respectively. The received signal here
can be expressed as $\mathbf{y}=g_{r}e^{j\phi_{r}}\mathbf{b}\left(\theta_{t},r_{t}\right)$.
Note that the optimization problem in \eqref{eq:GLRT} is equivalent
to 
\begin{equation}
\arg\max_{\theta,r}\left|\frac{1}{M}\sum_{\delta_{m}}e^{jk(r_{m}-r_{t,m})}\right|.
\end{equation}
Thus, we can denote the optimization problem as $\arg\max_{\theta,r}s(\theta,r)$.
By using the second-order approximation of $r_{m}=r+\delta_{m}d\cos\theta+\frac{\delta_{m}^{2}d^{2}\sin^{2}\theta}{2r}$,
we can obtain
\begin{equation}
\begin{aligned}s\left(\theta,r\right) & \approx\left|\frac{1}{M}\sum_{\delta_{m}}e^{j\delta_{m}\pi\left(\cos\theta-\cos\theta_{t}\right)+jk\delta_{m}^{2}d^{2}\left(\frac{\sin^{2}\theta}{2r}-\frac{\sin^{2}\theta_{t}}{2r_{t}}\right)}\right|.\end{aligned}
\label{eq:costFuncStan}
\end{equation}
In fact\textcolor{black}{, directly deriving the codebook requirements
from this complex 2D form remains analytically intractable. However,
we observe that the first term dominates the value of $s(\theta,r)$,
a property consistent with Taylor expansion theory. From a physical
perspective, the angle parameter $\theta$ exerts a primary influence
on the cost function. Once the codewords are closely aligned in angle,
the impact of the distance parameter $r$ becomes prominent. Guided
by this physical insight, we execute the aforementioned decoupled
approach, analyzing the angle and distance grid designs sequentially.}
\begin{figure}[t]
\centering{}\hspace{-0.8cm}\subfloat[]{
\centering{}\includegraphics[width=5cm]{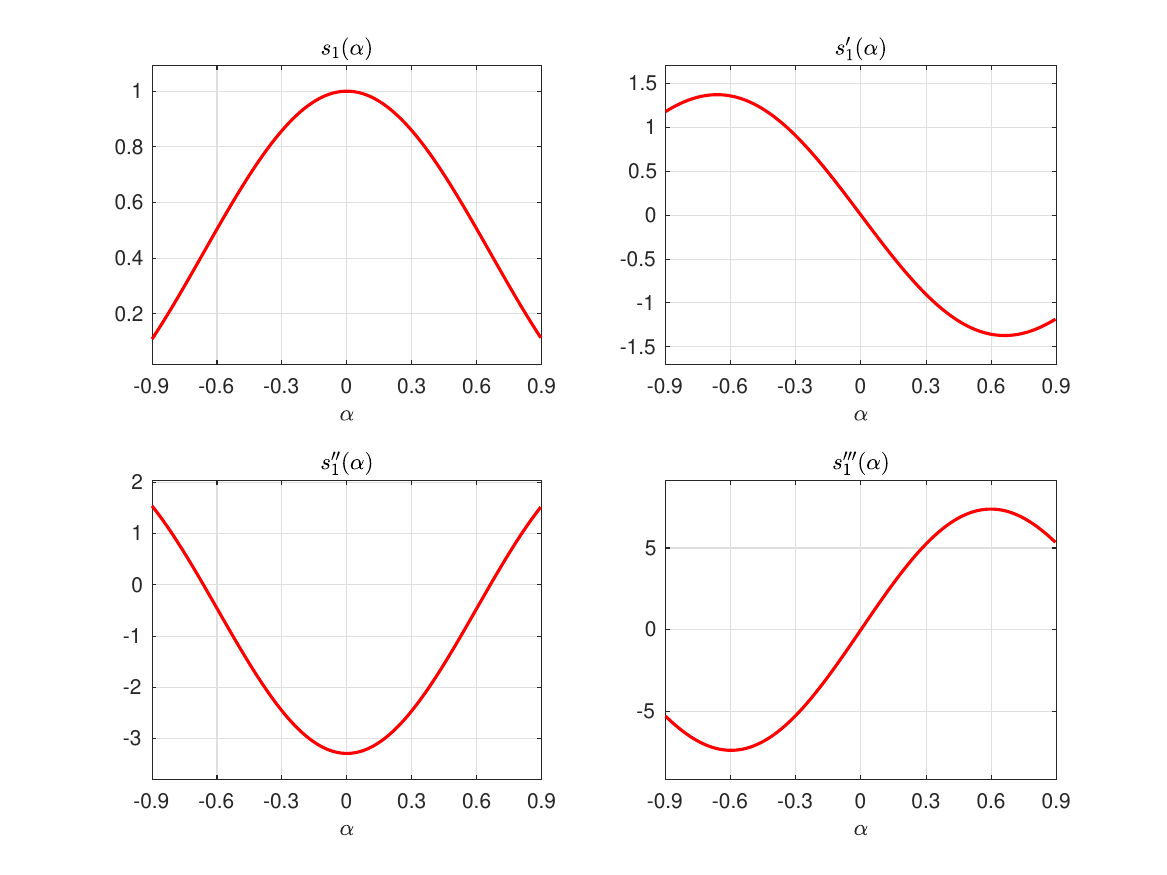}\hspace{-0.3cm}}\subfloat[]{\centering{}\includegraphics[width=5cm]{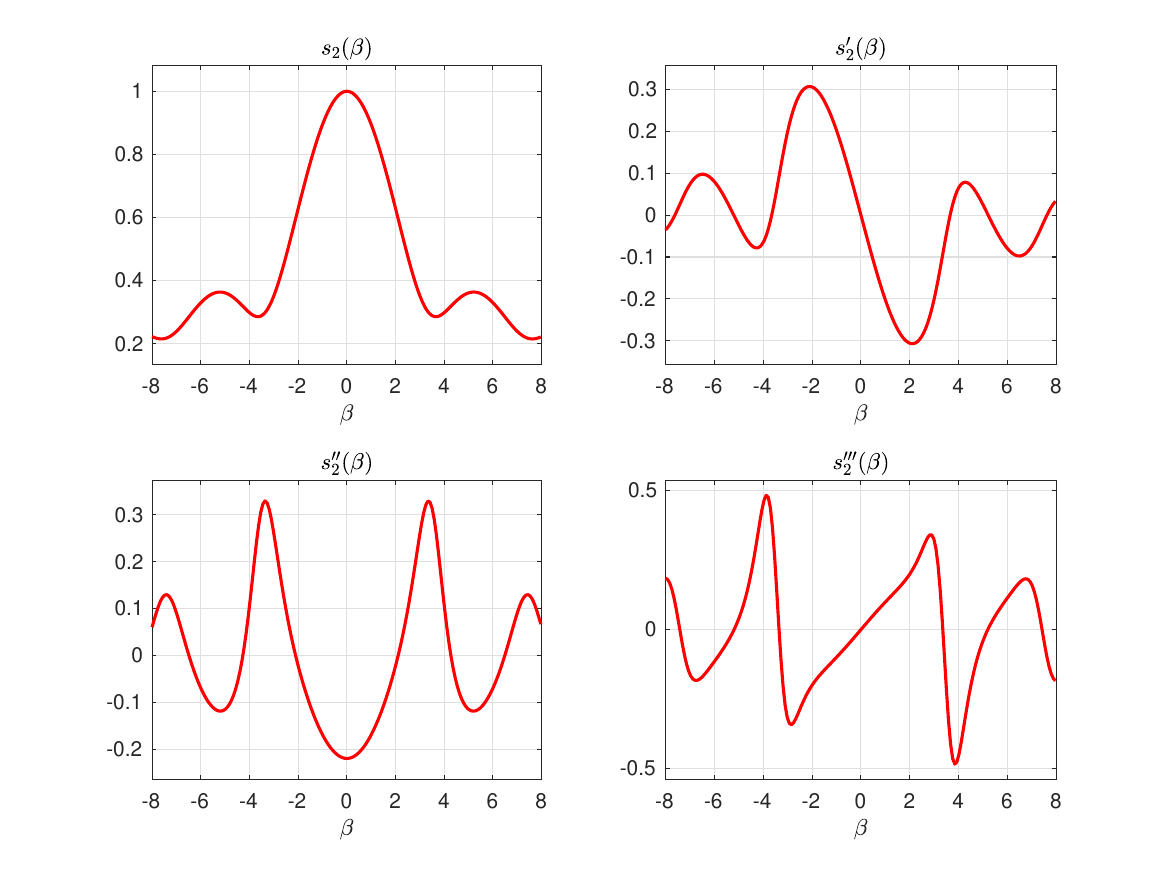}\hspace{-0.3cm}}\caption{(a) The first term $s_{1}\left(\alpha\right)$ and its derivatives.
(b) The second term $s_{2}\left(\beta\right)$ and its derivatives.
\label{fig: itemDerivative-both}}
\end{figure}

\subsection{\textcolor{black}{Angle Grid Design}}

When considering the codewords with different angles, we focus on
the contribution of the first term of \eqref{eq:costFuncStan}, denoted
as $s_{1}(\theta,r)=|\frac{1}{M}\sum_{\delta_{m}=-\frac{M-1}{2}}^{\frac{M-1}{2}}e^{j\delta_{m}\pi(\cos\theta-\cos\theta_{t})}|$.
When $M\gg1$, this term can be approximated as%
\begin{equation}
s_{1}\left(\theta,r\right)\approx\left|\frac{1}{\pi\alpha}\intop_{t=-\alpha}^{\alpha}e^{jt}dt\right|=s_{1}\left(\alpha\right)=\left|\frac{\sin\left(\pi\alpha\right)}{\pi\alpha}\right|,
\end{equation}
where $\alpha=\frac{1}{2}M(\cos\theta-\cos\theta_{t})$\textcolor{black}{.
From Fig. \ref{fig: itemDerivative-both}(a), which illustrates the
function $s_{1}\left(\alpha\right)$ and its }derivatives around \textcolor{black}{the
true solution} $\alpha_{t}=0$, we can observe that the first two
conditions in \textit{Claim 1} are satisfied when $\epsilon_{0}<0.66$.
After necessary mathematical calculations, we can obtain $\epsilon_{0}<0.45$
is the sufficient condition of the third condition in \textit{Claim
1}. Thus, the maximum acceptable grid spacing $\Delta_{\alpha}$ of
$\alpha$ is about $0.9$. In general, for different codewords in
the codebook, the cosine of angle $\cos\theta$ should be uniformly
sampled as
\begin{equation}
\cos\theta_{n_{\theta}}=\frac{2n_{\theta}\Delta_{\alpha}-M+1}{M},\ n_{\theta}=0,1,\ldots,\left\lfloor \frac{M}{\Delta_{\alpha}}\right\rfloor -1.
\end{equation}

\subsection{\textcolor{black}{Distance Grid Design}}

When considering the codewords with the same angle, the significance
of the second item in \eqref{eq:costFuncStan} becomes prominent.
We denote it as $s_{2}(\theta,r)=|\frac{1}{M}\sum_{\delta_{m}}e^{jk\delta_{m}^{2}d^{2}(\frac{\sin^{2}\theta}{2r}-\frac{\sin^{2}\theta_{t}}{2r_{t}})}|$.
Without loss of generality, we assume that there is a \textcolor{black}{set} of codewords
can well describe the true angle, i.e., $\theta=\theta_{t}$ and $s_{2}(\theta,r)=|\frac{1}{M}\sum_{\delta_{m}}e^{j\frac{k\delta_{m}^{2}d^{2}\sin^{2}\theta}{2}(\frac{1}{r}-\frac{1}{r_{t}})}|$.
In practice, this is equivalent to designing the distance parameters
for codewords with the same angle in the near-field codebook. We introduce
the Fresnel functions to approximate the intricate function as \cite{cui2022channel}
\begin{equation}
s_{2}\left(\theta,r\right)\approx s_{2}\left(\beta\right)=\begin{cases}
\left|\frac{C(\sqrt{\beta})+jS(\sqrt{\beta})}{\sqrt{\beta}}\right| & \beta>0\\
\left|\frac{C(\sqrt{-\beta})+jS(\sqrt{-\beta})}{\sqrt{-\beta}}\right| & \beta<0
\end{cases},\label{eq:2item}
\end{equation}
where $\beta=\frac{M^{2}d^{2}\sin^{2}\theta}{2\lambda}(\frac{1}{r}-\frac{1}{r_{t}})$.
The Fresnel sine and cosine integral \textcolor{black}{functions} are $S(x)=\int_{0}^{x}\sin(\frac{\pi}{2}t^{2})dt$
and $C(x)=\int_{0}^{x}\cos(\frac{\pi}{2}t^{2})dt$, respectively.
\textcolor{black}{From Fig. \ref{fig: itemDerivative-both}(b), which
illustrates the function $s_{2}\left(\beta\right)$ and its }derivatives
around\textcolor{black}{{} the true solution $\beta_{t}=0$}, the
first two conditions in \textit{Claim 1} can be met when $\epsilon_{0}<2.08$.%
{} It follows that the third condition in \textit{Claim 1} will be
satisfied when $\epsilon_{0}<1.49$ with some mathematical manipulations,
and the acceptable grid spacing $\Delta_{\beta}$ of $\beta$ is $\Delta_{\beta}<2.98$.
Therefore, for codewords with the same angle parameter $\theta$
in the codebook, the reciprocal of the distance parameter $\frac{1}{r}$
should be uniformly sampled as
\begin{equation}
\frac{1}{r_{n_{r}}}=\frac{2n_{r}\lambda\Delta_{\beta}}{M^{2}d^{2}\sin^{2}\theta},\ n_{r}=1,2,3,\ldots.
\end{equation}
Note that the range of \ensuremath{r} should be limited to $(1.2D,r_{R}]$.
If $r>r_{R}$ and $n_{r}=1$, we select the codeword with $(\theta,r_{R})$
as the only codeword in the codebook with the angle $\theta$. 

\section{Near-Field Soft Cooperative Localization\label{sec:SCL}}

In this section, we present a novel cooperative localization algorithm
based on Gaussian fusion. The GFCL algorithm utilizes the propagation
parameter estimates provided by the VNNCE algorithm to obtain soft
user positions for each BS. It then employs an appropriate strategy
to fuse these soft positions, thereby enhancing the accuracy of the
localization process. Furthermore, we propose a joint architecture
that mutually enhances both channel estimation and cooperative localization. 

\subsection{Soft Relative Position }

According to the spatial model \eqref{eq:SpaceModel}, the relative
position between the user or scatterer and each BS (the superscript
of BS index $i$ and the subscript of path index $l$ are omitted)
can be represented as
\begin{equation}
\left[\begin{array}{c}
x_{\mathrm{r}}\\
y_{\mathrm{r}}
\end{array}\right]=r\left[\begin{array}{c}
\cos\left(\theta+\omega\right)\\
\sin\left(\theta+\omega\right)
\end{array}\right].\label{eq:relativePosition}
\end{equation}
Furthermore, the expressions for the propagation parameters can be
easily obtained using the relative position parameters. They are given
by 
\begin{equation}
r=\sqrt{x_{r}^{2}+y_{r}^{2}},\label{eq:rTransform}
\end{equation}
\begin{equation}
\theta=\operatorname{arccot}\left(\frac{x_{\mathrm{r}}\cos(\omega)+y_{\mathrm{r}}\sin(\omega)}{y_{\mathrm{r}}\cos(\omega)-x_{\mathrm{r}}\sin(\omega)}\right).\label{eq:thetaTransform}
\end{equation}
With soft channel estimation, we can obtain the Gaussian representation
of the propagation parameters. In fact, the propagation parameters
are position parameters in a polar coordinate system with the BS as
the origin. However, for cooperative localization, it is more suitable
to work with the relative position parameters, which are the position
parameters in a unified Cartesian coordinate system. Similar to the
propagation parameters, the relative position parameters can also
be expressed in Gaussian form, given by

\textcolor{black}{
\begin{equation}
\boldsymbol{\nu}\sim\mathcal{N}\left(\hat{\boldsymbol{\nu}},\mathbf{\hat{V}_{\boldsymbol{\nu}}}\right),
\end{equation}
}where $\hat{\boldsymbol{\nu}}=(\hat{x_{r}},\hat{y_{r}})^{T}$ is
the relative position estimate, and $\mathbf{\hat{V}_{\boldsymbol{\nu}}}=\left[\begin{array}{cc}
\hat{v}_{x_{r}x_{r}} & \hat{v}_{x_{r}y_{r}}\\
\hat{v}_{x_{r}y_{r}} & \hat{v}_{y_{r}y_{r}}
\end{array}\right]$ represents its confidence level. The relative parameter estimate
can be obtained by substituting the propagation parameter estimate
into \eqref{eq:relativePosition}, while determining the confidence
level may present some challenges.
\begin{claim}
Let $y=f(u)$ and $u=g(x)$. If $h(x)=f(g(x))$ for every $x$, then
the second derivative, using \textit{Faà di Bruno's formula}, can
be expressed as 
\begin{equation}
\frac{d^{2}y}{dx^{2}}=\frac{d^{2}y}{du^{2}}\left(\frac{du}{dx}\right)^{2}+\frac{dy}{du}\frac{d^{2}u}{dx^{2}}.\label{eq:chainRule}
\end{equation}
\end{claim}
To determine the \textcolor{black}{confidence level matrix $\mathbf{\hat{V}_{\boldsymbol{\nu}}}$},
we need to compute the \textcolor{black}{Hessian matrix} of the optimization
function \eqref{eq:optFunc}, which is given by\textcolor{black}{
\begin{equation}
\nabla^{2}f(x_{r},y_{r})=\left[\begin{array}{cc}
\frac{\partial^{2}f(\mathbf{\boldsymbol{\mu}})}{\partial x_{r}^{2}} & \frac{\partial^{2}f(\mathbf{\boldsymbol{\mu}})}{\partial x_{r}\partial y_{r}}\\
\frac{\partial^{2}f(\mathbf{\boldsymbol{\mu}})}{\partial x_{r}\partial y_{r}} & \frac{\partial^{2}f(\mathbf{\boldsymbol{\mu}})}{\partial y_{r}^{2}}
\end{array}\right].
\end{equation}
}However, directly substituting \eqref{eq:rTransform} and \eqref{eq:thetaTransform}
into \eqref{eq:optFunc} is complex. To simplify this process, we
can utilize \textit{Claim 2} to obtain the \textcolor{black}{elements
of the Hessian matrix} as\begin{subequations} \label{secondDerivative2}
\begin{align}
\frac{\partial^{2}f(\mathbf{\boldsymbol{\mu}})}{\partial x_{r}^{2}} & =\left(\frac{\partial\theta}{\partial x_{r}}\right)^{2}\frac{\partial^{2}f(\mathbf{\boldsymbol{\mu}})}{\partial\theta^{2}}+\frac{\partial^{2}\theta}{\partial x_{r}^{2}}\frac{\partial f(\mathbf{\boldsymbol{\mu}})}{\partial\theta}\nonumber \\
 & +\left(\frac{\partial r}{\partial x_{r}}\right)^{2}\frac{\partial^{2}f(\mathbf{\boldsymbol{\mu}})}{\partial r^{2}}+\frac{\partial^{2}r}{\partial x_{r}^{2}}\frac{\partial f(\mathbf{\boldsymbol{\mu}})}{\partial r},
\end{align}
\begin{align}
\frac{\partial^{2}f(\mathbf{\boldsymbol{\mu}})}{\partial y_{r}^{2}} & =\left(\frac{\partial\theta}{\partial y_{r}}\right)^{2}\frac{\partial^{2}f(\mathbf{\boldsymbol{\mu}})}{\partial\theta^{2}}+\frac{\partial^{2}\theta}{\partial y_{r}^{2}}\frac{\partial f(\mathbf{\boldsymbol{\mu}})}{\partial\theta}\nonumber \\
 & +\left(\frac{\partial r}{\partial y_{r}}\right)^{2}\frac{\partial^{2}f(\mathbf{\boldsymbol{\mu}})}{\partial r^{2}}+\frac{\partial^{2}r}{\partial y_{r}^{2}}\frac{\partial f(\mathbf{\boldsymbol{\mu}})}{\partial r},
\end{align}
\begin{align}
\frac{\partial^{2}f(\mathbf{\boldsymbol{\mu}})}{\partial x_{r}\partial y_{r}} & =\frac{\partial\theta}{\partial x_{r}}\frac{\partial\theta}{\partial y_{r}}\frac{\partial^{2}f(\mathbf{\boldsymbol{\mu}})}{\partial\theta^{2}}+\frac{\partial^{2}\theta}{\partial x_{r}\partial y_{r}}\frac{\partial f(\mathbf{\boldsymbol{\mu}})}{\partial\theta}\nonumber \\
 & +\frac{\partial r}{\partial x_{r}}\frac{\partial r}{\partial y_{r}}\frac{\partial^{2}f(\mathbf{\boldsymbol{\mu}})}{\partial r^{2}}+\frac{\partial^{2}r}{\partial x_{r}\partial y_{r}}\frac{\partial f(\mathbf{\boldsymbol{\mu}})}{\partial r}.
\end{align}
\end{subequations} The \textcolor{black}{components of the gradient
vector and the elements of the Hessian matrix with respect to} $(\theta,r)$
are provided in \eqref{firstDerivative} and \eqref{secondDerivative},
respectively. The coefficients associated with these derivatives are
given by \begin{subequations} \label{coefficients} 
\begin{equation}
\frac{\partial\theta}{\partial x_{r}}=-\frac{y_{r}}{x_{r}^{2}+y_{r}^{2}},\ \frac{\partial\theta}{\partial y_{r}}=\frac{x_{r}}{x_{r}^{2}+y_{r}^{2}},
\end{equation}
\begin{equation}
\frac{\partial r}{\partial x_{r}}=\frac{x_{r}}{\sqrt{x_{r}^{2}+y_{r}^{2}}},\ \frac{\partial r}{\partial y_{r}}=\frac{y_{r}}{\sqrt{x_{r}^{2}+y_{r}^{2}}},
\end{equation}
\begin{equation}
\frac{\partial^{2}\theta}{\partial x_{r}^{2}}=\frac{2x_{r}y_{r}}{(x_{r}^{2}+y_{r}^{2})^{2}},\ \frac{\partial^{2}\theta}{\partial y_{r}^{2}}=\frac{2x_{r}y_{r}}{(x_{r}^{2}+y_{r}^{2})^{2}},
\end{equation}
\begin{equation}
\frac{\partial^{2}\theta}{\partial x_{r}\partial y_{r}}=\frac{y_{r}^{2}-x_{r}^{2}}{(x_{r}^{2}+y_{r}^{2})^{2}},\ \frac{\partial^{2}r}{\partial x_{r}^{2}}=\frac{y_{r}^{2}}{\left(\sqrt{x_{r}^{2}+y_{r}^{2}}\right)^{3}},
\end{equation}
\begin{equation}
\frac{\partial^{2}r}{\partial y_{r}^{2}}=\frac{x_{r}^{2}}{\left(\sqrt{x_{r}^{2}+y_{r}^{2}}\right)^{3}},\ \frac{\partial^{2}r}{\partial x_{r}\partial y_{r}}=-\frac{x_{r}y_{r}}{\left(\sqrt{x_{r}^{2}+y_{r}^{2}}\right)^{3}}.
\end{equation}
\end{subequations} Then, the confidence level can be calculated using\textcolor{black}{
\begin{equation}
\hat{V}_{\nu}=-[\nabla^{2}f(x_{r},y_{r})]^{-1}.\label{eq:varianceCalculation}
\end{equation}
}Therefore, we can obtain the soft relative positions from each path.
Additionally, by combining the known BS position $\mathbf{x}_{b}$,
we can determine the soft user or scatterer position using \eqref{eq:SpaceModel}. 

\subsection{Soft Position Fusion}

Because all of the soft positions are defined as Gaussian variables,
we can utilize the rule of multiple Gaussian products, as presented
in \textit{Claim 3}, to fuse these positions.
\begin{claim}
Let 
\[
f_{N}\left(\mathbf{x};\mathbf{m}_{1},\mathbf{V}_{1}\right),f_{N}\left(\mathbf{x};\mathbf{m}_{2},\mathbf{V}_{2}\right),\ldots,f_{N}\left(\mathbf{x};\mathbf{m}_{K},\mathbf{V}_{K}\right)
\]
be $K$ Gaussian PDFs with mean vectors $\mathbf{m}_{1},\mathbf{m}_{2},\ldots,\mathbf{m}_{K}$
and covariance \textcolor{black}{matrices} $\mathbf{V}_{1},\mathbf{V}_{2},\ldots,\mathbf{V}_{K}$,
respectively. Then, the product of these Gaussian PDFs is a scaled
Gaussian. It is given by
\begin{equation}
\prod_{k=1}^{K}f_{N}\left(\mathbf{x};\mathbf{m}_{k},\mathbf{V}_{k}\right)=Af_{N}\left(\mathbf{x};\mathbf{m}_{u},\mathbf{V}_{u}\right),\label{eq:GaussianFusion}
\end{equation}
where $\mathbf{V}_{u}=(\sum_{k=1}^{K}\mathbf{V}_{k}^{-1})^{-1}$,
$\mathbf{m}_{u}=\mathbf{V}_{u}\sum_{k=1}^{K}\mathbf{V}_{k}^{-1}\mathbf{m}_{k}$
and $A$ is a constant.

However, it is important to note that the rule of multiple Gaussian
products holds true only when the soft positions come from the same
target, such as the user. If the soft positions come from different
targets, the fusion process will be destructive. Therefore, it becomes
necessary to design a strategy to associate these soft positions appropriately.
To determine the association of soft positions, we define the cost
of a Gaussian \textcolor{black}{random variable $x\sim\mathcal{N}(m,V)$
as its} the mean squared error, which is given by 
\begin{equation}
\mathrm{E}_{\mathbf{x}}\left\{ \left\Vert \mathbf{x}-\mathbf{m}_{\mathrm{}}\right\Vert _{2}^{2}\right\} =\operatorname{tr}(\mathbf{V}).\label{eq:cost}
\end{equation}
In the localization problem, we consider a more accurate estimate
to have a lower cost. 

First, our objective is to find the most probable LoS path among all
the paths for each BS. UM-MIMO systems usually rely on high-frequency
communication, which offer the advantage of reducing the size of antenna
elements, allowing for a larger number of antennas to be placed within
a limited area \cite{rappaport2019wireless}. In high-frequency communication
systems, the signal energy is primarily concentrated on the
LoS path. Therefore, the LoS path will have the largest cost according
to \eqref{eq:costFunc}, and the soft position calculated from the
LoS path will have the least cost according to \eqref{eq:cost}. \textcolor{black}{This suggests that, if present, the LoS path is the first path extracted by VNNCE.} In practical environments, the LoS path may be blocked
by barriers, resulting in the soft scatterer position having the least
cost. Additionally, in situations with poor wireless channel conditions,
false estimates of propagation parameters may occur, leading to the
soft user position being far from the actual position.

To address these challenges, we employ a consistency strategy \cite{yang2022soft}
to determine whether each soft position contributes constructively
or destructively to the final user localization. Given two soft positions
\textcolor{black}{$\boldsymbol{\nu}_{1}\sim\mathcal{N}\left(\hat{\boldsymbol{\nu}}_{1},\mathbf{\hat{V}}_{\boldsymbol{\nu}_{1}}\right)$}
and \textcolor{black}{$\boldsymbol{\nu}_{2}\sim\mathcal{N}\left(\hat{\boldsymbol{\nu}}_{2},\mathbf{\hat{V}}_{\boldsymbol{\nu}_{2}}\right)$},
we calculate the consistency indicator as:
\begin{equation}
\eta=\begin{cases}
1, & (\hat{\boldsymbol{\nu}}_{1}-\hat{\boldsymbol{\nu}}_{2})^{T}(\mathbf{\hat{V}}_{\boldsymbol{\nu}_{1}}+\mathbf{\hat{V}}_{\boldsymbol{\nu}_{2}})^{-1}(\hat{\boldsymbol{\nu}}_{1}-\hat{\boldsymbol{\nu}}_{2})<\zeta^{2}\\
0, & \text{ otherwise }
\end{cases},\label{eq:consistencyIndicator}
\end{equation}
where $\zeta$ , a tunable parameter, is the consistency threshold.
\textcolor{black}{Mathematically, the term $(\hat{\boldsymbol{\nu}}_{1}-\hat{\boldsymbol{\nu}}_{2})^{T}(\mathbf{\hat{V}}_{\boldsymbol{\nu}_{1}}+\mathbf{\hat{V}}_{\boldsymbol{\nu}_{2}})^{-1}(\hat{\boldsymbol{\nu}}_{1}-\hat{\boldsymbol{\nu}}_{2})<\zeta^{2}$
represents the squared Mahalanobis distance between the two Gaussian
distributions. This threshold acts as a statistical gate: if
the Mahalanobis distance is less than $\zeta^{2}$ (i.e., $\eta=1$),
the two soft positions are deemed to represent the identical physical
target and are constructively fused. Conversely, if the distance exceeds
the threshold, the position is rejected as an outlier (e.g., a severe
NLoS component misidentified as a LoS path), thereby preventing destructive
interference during the fusion process.}
\begin{algorithm}[t]
\caption{Proposed GFCL Algorithm. \label{alg:GFCL}}

\textbf{Input:} Propagation parameters \textcolor{black}{$\{\boldsymbol{\mu}_{l}^{i}\sim\mathcal{N}\left(\hat{\boldsymbol{\mu}}_{l}^{i},\mathbf{\hat{V}_{\boldsymbol{\mu}_{l}^{i}}}\right):i=1,2,\ldots,I,\ l=1,2,\ldots,L_{i}'\}$},
Consistency threshold $\zeta$;

\begin{algorithmic}[1]

\Statex // Soft User Position Inference

\State Infer all of the soft relative position parameters \textcolor{black}{$\{\boldsymbol{\nu}_{l}^{i}\sim\mathcal{N}\left(\hat{\boldsymbol{\nu}}_{l}^{i},\mathbf{\hat{V}_{\boldsymbol{\nu}_{l}^{i}}}\right):i=1,2,\ldots,I,\ l=1,2,\ldots,L_{i}'\}$
}by \textcolor{black}{\eqref{eq:relativePosition}} and \textcolor{black}{\eqref{eq:varianceCalculation}};

\State Select the soft relative position $\boldsymbol{\nu}^{i}$\textbf{
}with the least cost for each BS $i$, and consider it as the soft
relative position between the user and BS $i$:
\[
\boldsymbol{\nu}^{i}\leftarrow\arg\min_{\boldsymbol{\nu}_{l}^{i}}\operatorname{tr}\left(\mathbf{\hat{V}}_{\boldsymbol{\nu}_{l}^{i}}\right);
\]

\State Substitute $\{\boldsymbol{\nu}^{i}:i=1,2,\ldots,I\}$ into
\eqref{eq:SpaceModel}, then we obtain the soft user positions $\{\boldsymbol{\nu}_{u}^{i}:i=1,2,\ldots,I\}$;

\Statex // Soft User Position Association

\State Reindex the $I$ soft user positions in \textcolor{black}{ascending} order of cost:
\[
\operatorname{tr}\left(\mathbf{\hat{V}}_{\boldsymbol{\nu}_{u}^{1}}\right)<\operatorname{tr}\left(\mathbf{\hat{V}}_{\boldsymbol{\nu}_{u}^{2}}\right)<\ldots<\operatorname{tr}\left(\mathbf{\mathbf{\hat{V}}_{\boldsymbol{\nu}_{u}^{I}}}\right);
\]
\State Select the first soft user position $\boldsymbol{\nu}_{u}^{1}$
with least cost as the reference soft user position, and set $\eta^{i}=1$;

\State Calculate the consistency indicators $\{\eta^{i}:i=2,\ldots,I\}$
for $\{\boldsymbol{\nu}_{u}^{i}:i=2,\ldots,I\}$ by \eqref{eq:consistencyIndicator};

\Statex // Soft User Position Fusion

\State Fuse all soft user positions with $\eta^{i}=1$ by using \eqref{eq:GaussianFusion}
to obtain the user position estimate $\hat{\mathbf{x}}_{u}$:
\textcolor{black}{\[
\hat{\mathbf{x}}_{u}\leftarrow\sum_{i=1,\ \eta^{i}=1}^{I}\mathbf{\hat{V}}^{-1}_{\boldsymbol{\nu}_{u}^{i}}\hat{\boldsymbol{\nu}}_{u}^{i}\left(\sum_{i=1,\ \eta^{i}=1}^{I}\mathbf{\hat{V}}^{-1}_{\boldsymbol{\nu}_{u}^{i}}\right)^{-1};
\]}

\end{algorithmic}

\textbf{Output:} User position estimate $\hat{\mathbf{x}}_{u}$.
\end{algorithm}

\begin{figure*}[t]
\begin{centering}
\includegraphics[width=12cm]{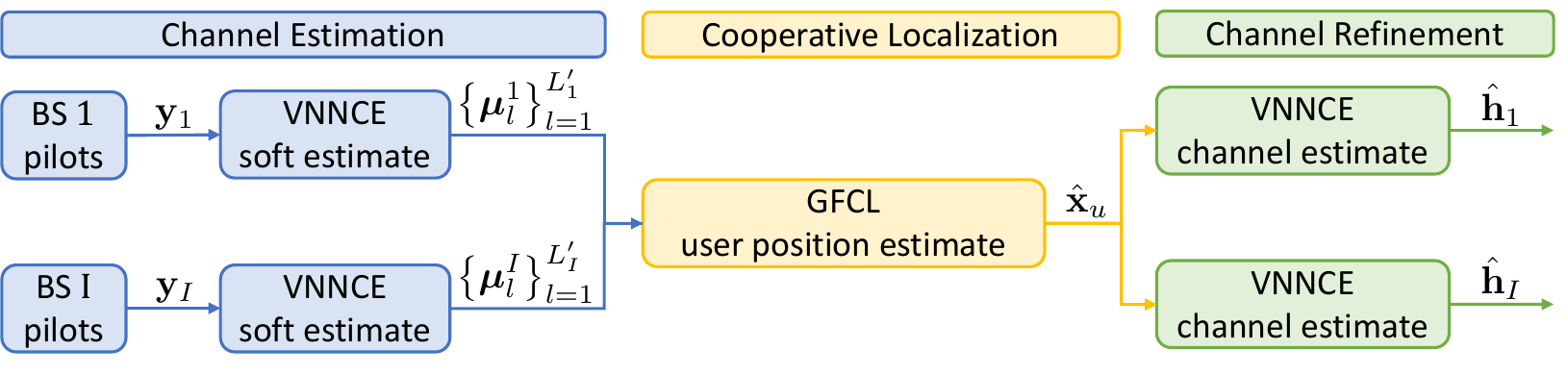}
\par\end{centering}
\caption{\textcolor{black}{Schematic of the proposed Joint Channel Estimation and Cooperative Localization architecture.
\label{fig:joint_vnnce_gfcl}}}
\end{figure*}

The GFCL algorithm\footnote{\textcolor{black}{To illustrate the consistency check in GFCL, consider a three-BS scenario. After cost-based sorting, the estimate from BS 1 has the smallest uncertainty trace and is selected as the initial reference, i.e., $\eta^1=1$. The estimate from BS 2 is then tested against this reference using the Mahalanobis distance. If the distance is within the threshold $\zeta$, BS 2 passes the consistency check and is accepted, yielding $\eta^2=1$. It then participates in constructive fusion, where the accepted estimates are combined using inverse-covariance weights. By contrast, BS 3 may provide a biased estimate, for example under severe NLoS propagation. When its Mahalanobis distance exceeds $\zeta$, the corresponding indicator is set to $\eta^3=0$. This leads to destructive fusion, in which the inconsistent estimate is removed from the fusion set and therefore does not affect the final user position estimate.}}, outlined in \textbf{Algorithm \ref{alg:GFCL}},
can be primarily divided into three steps, each serving the following
purposes: 
\end{claim}
\begin{enumerate}
\item \textit{Soft User Position Inference}: Steps 1-3 focus on determining
the most probable soft user position for fusion, as discussed earlier.
\item \textit{Soft User Position Association}: Steps 4-5 aim to extract
the most accurate soft user position as the reference position. Then,
in Step 6, the remaining soft user positions are associated with the
reference position. This association process helps assess whether
each soft user position contributes constructively or destructively
to the final user position estimate. By considering these associations,
we can evaluate the impact of each soft user position on the accuracy
and reliability of the cooperative localization result.
\item \textit{Soft User Position Fusion}: In Step 7, all the constructively
associated soft user positions are fused to obtain the final user
position estimate. Notably, the cost of the Gaussian fusion result
is lower than the costs of the individual Gaussian elements prior
to fusion. This implies that the cooperative localization accuracy
improves after fusion, providing a higher level of precision compared
to the independent localization accuracy of a single BS. 
\end{enumerate}

\subsection{Joint Design of Channel Estimation and Localization}

In stark contrast to conventional approaches that treat channel estimation
and localization as \textcolor{black}{independently} executed tasks, our framework establishes
a tightly coupled, bidirectional joint design. The core innovation
of our work lies not merely in performing both functions, but in enabling
them to mutually enhance each other through iterative refinement,
thereby exploiting the full potential of spatial \textcolor{black}{DoFs}
in near-field UM-MIMO systems.

Specifically, the VNNCE algorithm leverages a novel near-field codebook
that explicitly captures the curvature of spherical wavefronts and
satisfies the local \textcolor{black}{concavity} conditions required by Newton-type methods,
thereby ensuring convergence to the true channel parameters. On the
other hand, the VNNCE algorithm goes beyond point estimation. It outputs
uncertainty-aware soft estimates of near-field propagation parameters,
along with statistically meaningful confidence levels. These soft
outputs serve as inputs to the GFCL
algorithm, enabling principled fusion of multi-BS measurements while
accounting for estimation reliability. This constitutes channel-estimation-aided cooperative localization. \textcolor{black}{The resulting user-position estimate is then fed back to the channel-estimation module.} The high-accuracy user position estimate obtained by GFCL is
fed back to the channel estimation module. By leveraging this refined
geometric prior, we reconstruct the LoS component of the channel and
fix its parameters. The VNNCE algorithm is then re-executed on the
residual signal to refine the estimation of NLoS paths or channel
gains. This yields cooperative localization enhanced channel estimation.

To realize this joint vision, we propose a unified architecture called
Joint Channel Estimation and Cooperative Localization. \textcolor{black}{As illustrated in Fig.~\ref{fig:joint_vnnce_gfcl}, its} workflow consists of three tightly integrated stages:
\begin{enumerate}
\item \textit{Channel Estimation}: Each BS independently runs the VNNCE
algorithm on its received pilot signal to estimate near-field channel
parameters, including angle, distance, and complex gain. The output
includes not only point estimates but also confidence levels that
quantify estimation uncertainty.
\item \textit{Cooperative Localization}: A central server collects the soft
parameter estimates from all BSs and applies the GFCL algorithm. For
each BS with a LoS path, it computes a local position estimate. These
estimates are then fused in a statistically principled manner to produce
a high-accuracy global user position. 
\item \textit{Channel Refinement}: The refined user position from Step 2\textcolor{black}{,
together with the known BS coordinates and array orientations,}
is used to analytically determine the LoS path parameters (AoA and
distance) for each BS \textcolor{black}{via standard geometry}. These
\textcolor{black}{geometrically computed} parameters are fixed in the
channel model, and VNNCE is re-applied to the residual signal to refine
the estimation of remaining unknowns, such as NLoS components or complex
gains. This feedback loop significantly improves overall channel reconstruction
accuracy.
\begin{table}[t]
\caption{Channel Estimation Key Simulation Parameters \label{tab:KSP}}

\centering{}%
\begin{tabular}{|l|l|}
\hline 
\textbf{Parameter} & \textbf{Value}\tabularnewline
\hline 
Antenna number of ULAs & $M=256$\tabularnewline
Wavelength of carrier signals & $\lambda_{c}=3\ \textrm{mm}$\tabularnewline
Number of paths & $L=3$\tabularnewline
Number of single refinement rounds & $R_{s}=5$\tabularnewline
Number of cyclic refinement rounds & $R_{c}=5$\tabularnewline
Grid spacing of angles & $\Delta_{\alpha}=0.5$\tabularnewline
Grid spacing of distances & $\Delta_{\beta}=1$\tabularnewline
\hline 
\end{tabular}
\end{table}
\begin{figure}[t]
\begin{centering}
\includegraphics[width=5cm]{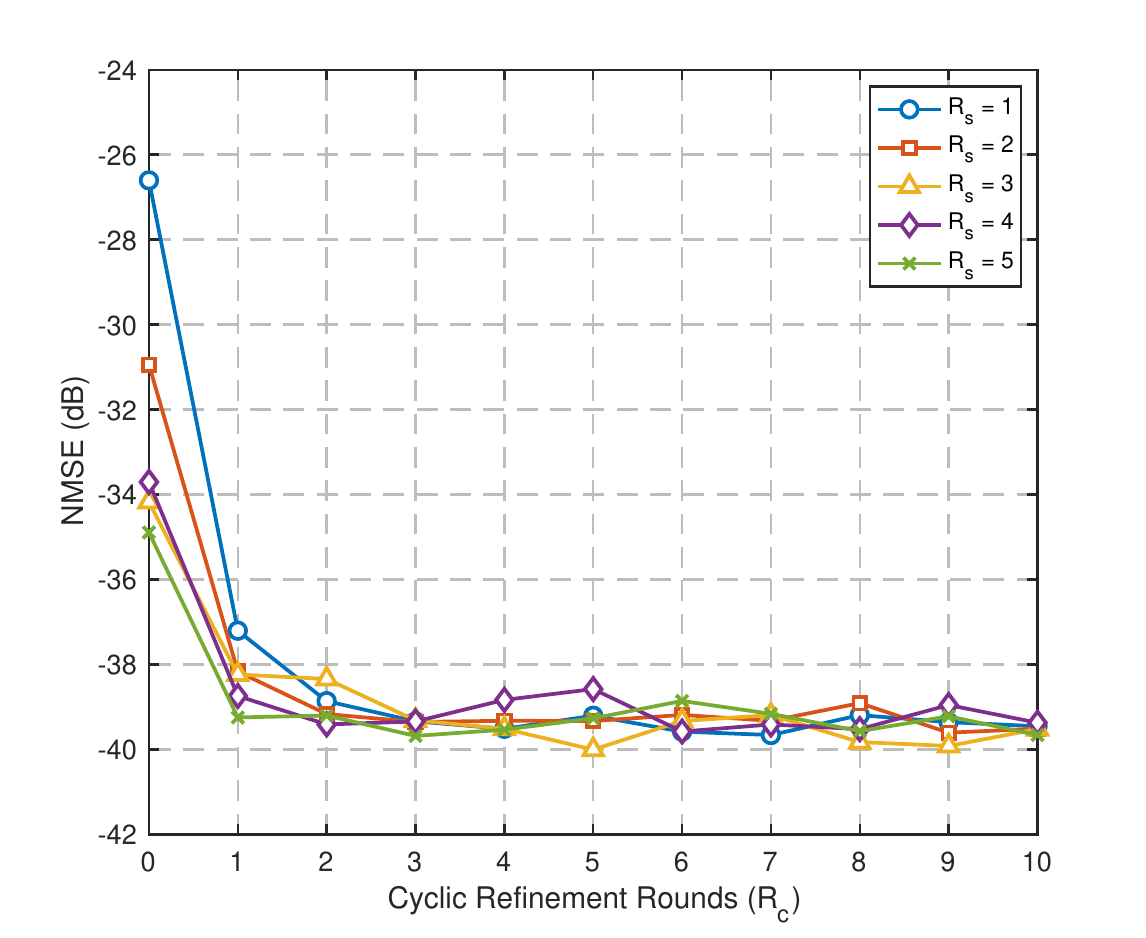}
\par\end{centering}
\caption{\textcolor{black}{Empirical convergence of the VNNCE algorithm at $\textrm{SNR}=25\,\textrm{dB}$.
\label{fig:convergence}}}
\end{figure}
\begin{figure}[t]
\begin{centering}
\includegraphics[width=5cm]{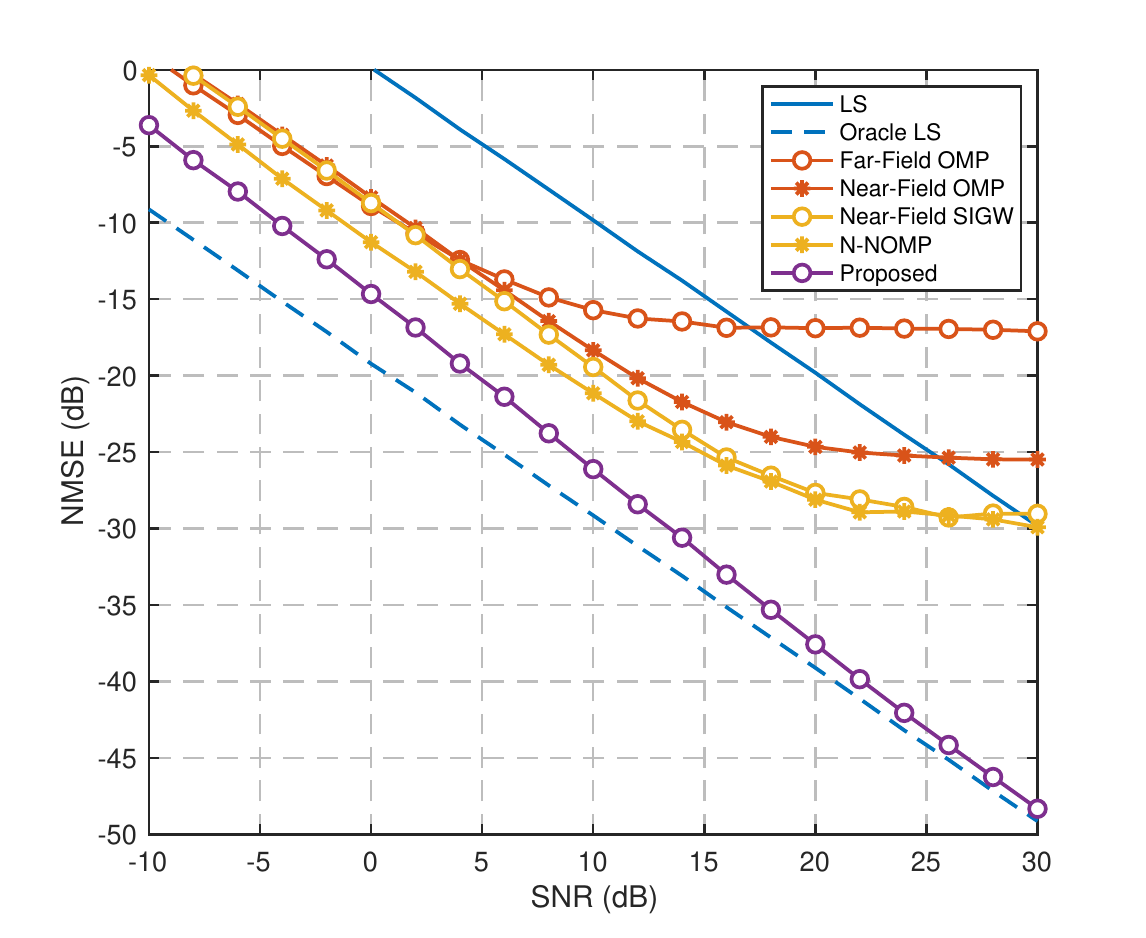}
\par\end{centering}
\caption{NMSE achieved by different algorithms versus SNR. \label{fig:NMSE}}
\end{figure}
\begin{table}[t]
\caption{\textcolor{black}{Average Execution Time Comparison \label{tab:RimTime}}}

\centering{}%
\begin{tabular}{|l|c|}
\hline 
\textbf{Algorithm} & \textbf{Average Execution Time (s)}\tabularnewline
\hline 
VNNCE & $0.0146$\tabularnewline
\hline 
P-SOMP \cite{cui2022channel} & $0.0053$\tabularnewline
\hline 
P-SIGW \cite{cui2022channel} & $0.1469$\tabularnewline
\hline 
N-NOMP \cite{lu2022near} & $0.0990$\tabularnewline
\hline 
\end{tabular}
\end{table}
\begin{table}[t]
\caption{Localization System Additional Key Simulation Parameters \label{tab:KSPA}}

\centering{}%
\begin{tabular}{|l|l|}
\hline 
\textbf{Parameter} & \textbf{Value}\tabularnewline
\hline 
Number of BSs & $I=4$\tabularnewline
Positions of BSs & $\mathbf{x}_{b}^{i}=\{(0,50),(20,50),(50,0),(50,20)\}$\tabularnewline
ULA rotations of BSs & $\omega^{i}=\{\pi,\pi,\pi/2,\pi/2\}$\tabularnewline
Position of user & $\mathbf{x}_{u}=(25,18)$\tabularnewline
Number of paths & $L^{i}=\{2,2,2,2\}$\tabularnewline
Noise variance & $\sigma^{2}=-110\ \textrm{dBm}$\tabularnewline
Consistency threshold & $\zeta=3.5$\tabularnewline
\hline 
\end{tabular}
\end{table}
\begin{figure}[t]
\centering{}\subfloat[]{
\centering{}\includegraphics[width=4.5cm]{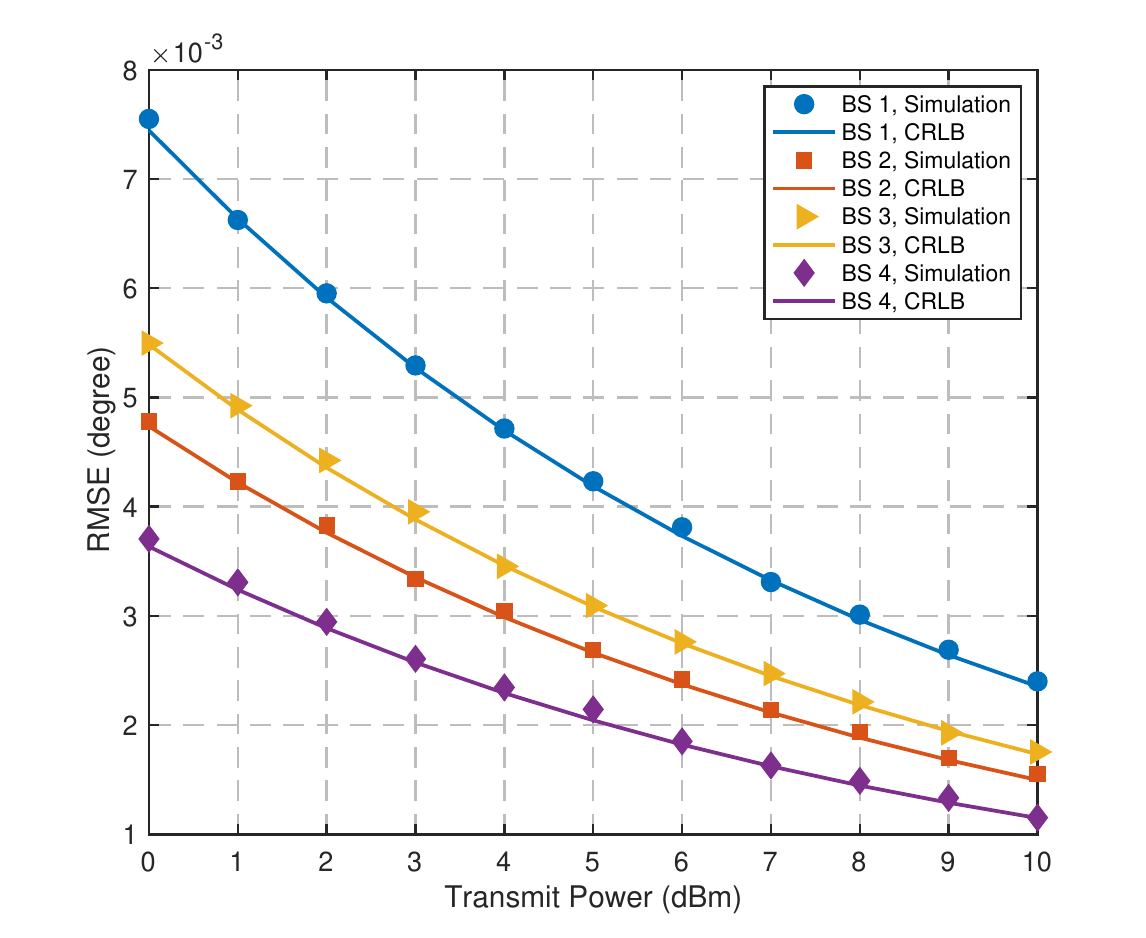}}\subfloat[]{\centering{}\includegraphics[width=4.5cm]{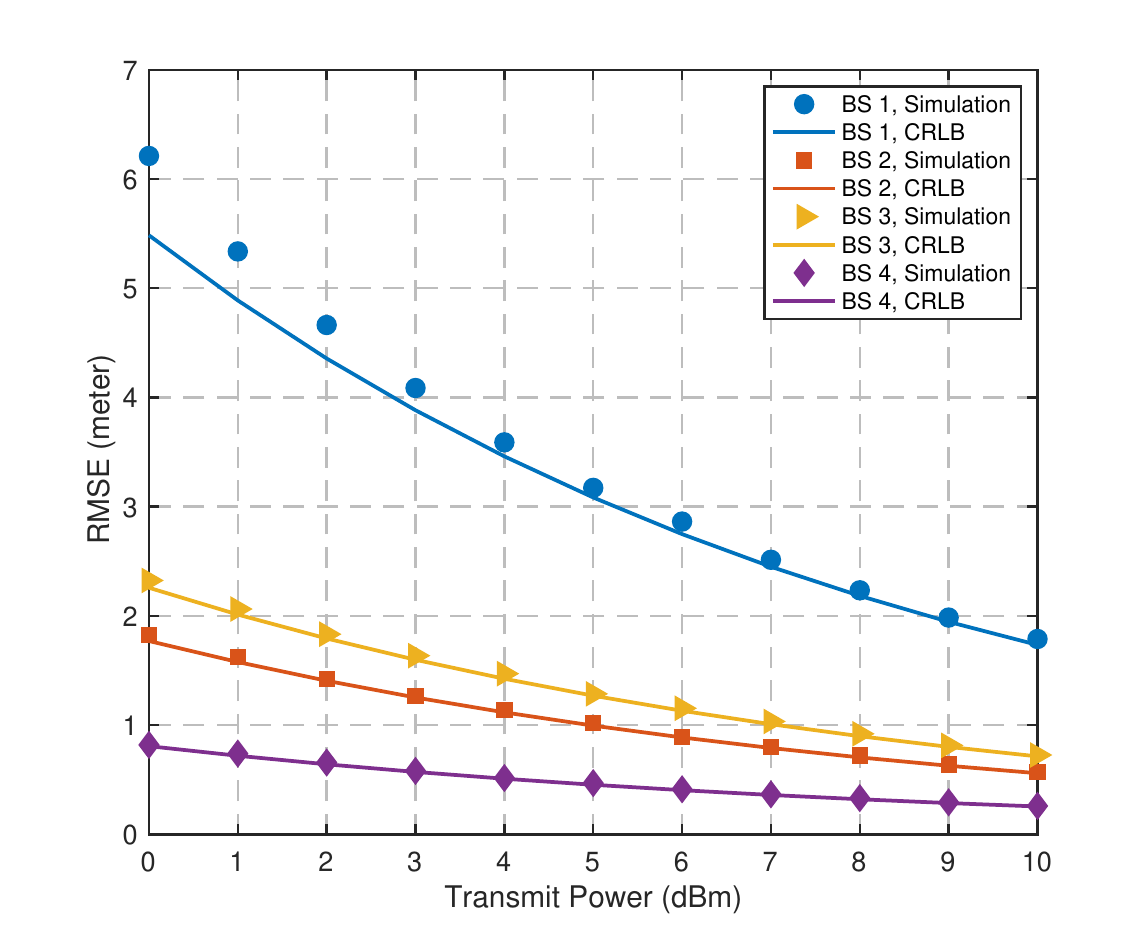}}\caption{(a) RMSE of the LoS path angle estimate $\hat{\theta}_{\textrm{LoS}}$.
(b) RMSE of the LoS path distance estimate $\hat{r}_{\textrm{LoS}}$.
\label{fig: angle-distance-Results}}
\end{figure}
\begin{figure}[t]
\centering{}\subfloat[]{
\centering{}\includegraphics[width=4.5cm]{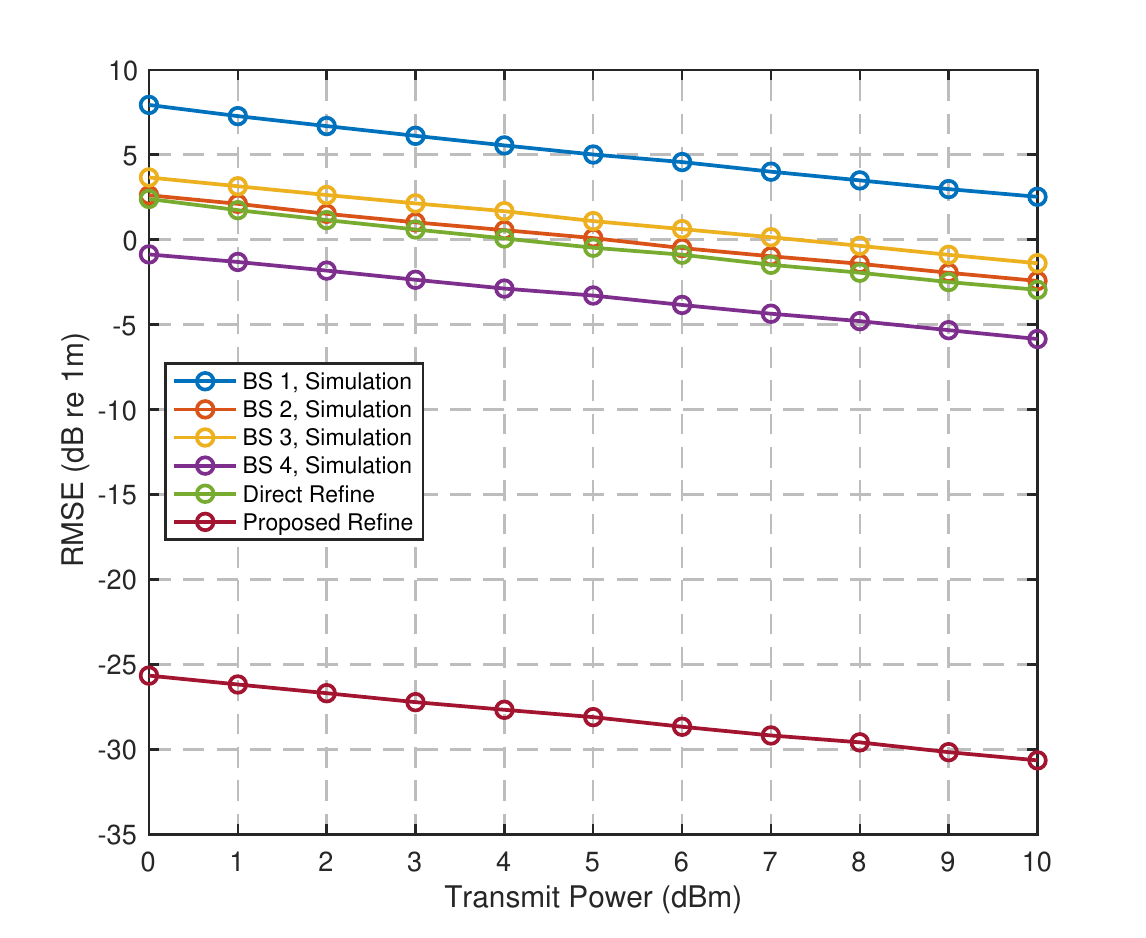}}\subfloat[]{\centering{}\includegraphics[width=4.5cm]{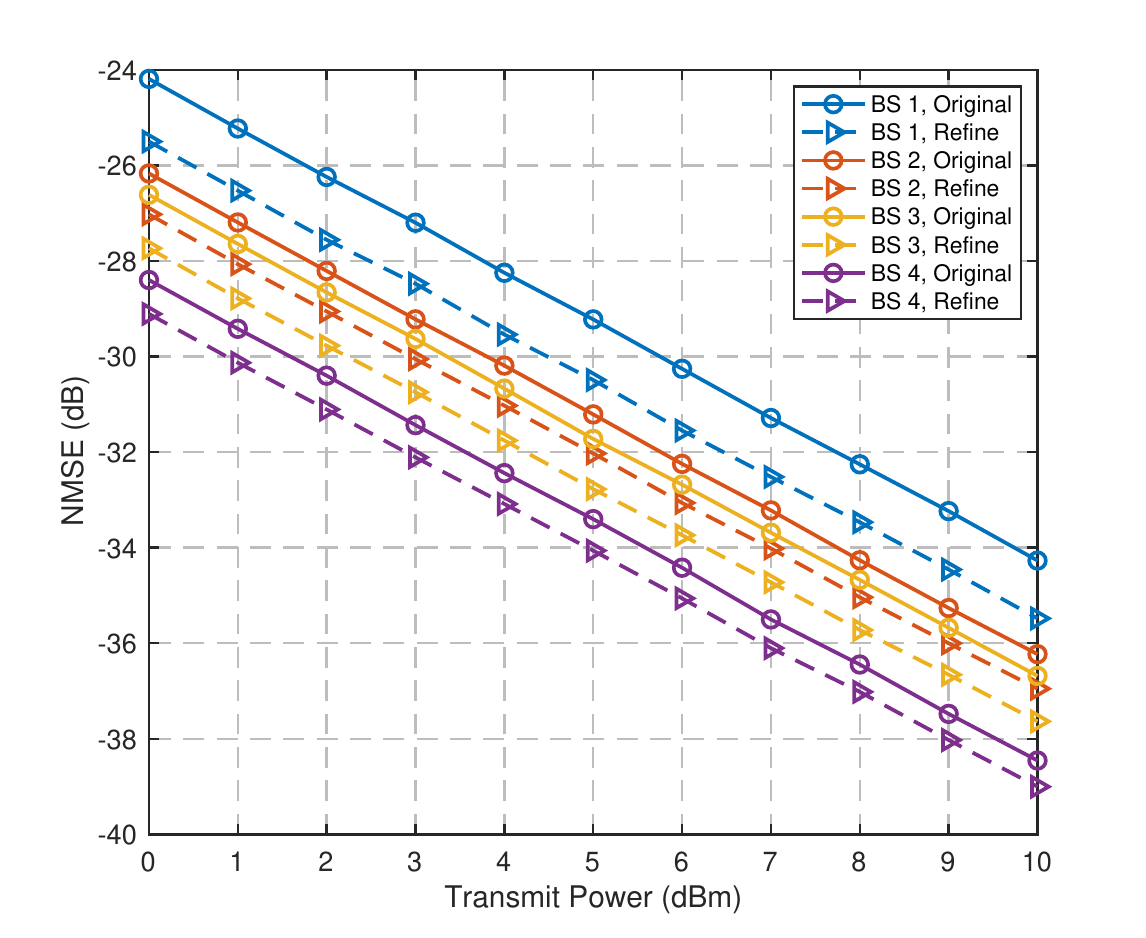}}\caption{(a) RMSE of soft cooperative localization through joint channel estimation
and localization. (b) NMSE of soft channel estimation through joint
channel estimation and localization. \label{fig: soft-results}}
\end{figure}
\end{enumerate}

\section{Simulation Results \label{sec:Simulation-Results}}

In this section, we begin by presenting simulation results to evaluate
the performance of the proposed VNNCE algorithm with respect to channel
estimation. We conduct these simulations in a typical near-field single-user
UM-MIMO system. 

We consider the \textcolor{black}{Normalized Mean Square Error (NMSE)}
as the performance metric for channel estimation, which is defined
as $\mathrm{NMSE}=\mathrm{E}_{\hat{\mathbf{H}}}\left\{ \frac{\|\mathbf{H}-\hat{\mathbf{H}}\|_{2}^{2}}{\|\mathbf{H}\|_{2}^{2}}\right\} $.
To conduct our simulations, we set the key parameters as listed in
Table \ref{tab:KSP}. For each path, the angle $\theta_{l}\in(0,\pi)$
and distance $r_{l}\in(1.2D,r_{R}]$ are randomly and uniformly generated
within the specified range. \textcolor{black}{Additionally, all the
presented statistical results are averaged over $5000$ independent
Monte Carlo channel realizations to ensure high reliability.} 

\textcolor{black}{Before evaluating the estimation accuracy against
the SNR, we first provide empirical evidence regarding the convergence
behavior of the proposed VNNCE framework. Fig. \ref{fig:convergence}
illustrates the channel estimation NMSE versus the number of cyclic
refinement rounds $R_{c}$ for varying single refinement steps $R_{s}\in\{1,2,3,4,5\}$
at a fixed SNR of $25\,\textrm{dB}$, with other parameters matched
to Table \ref{tab:KSP}. The results show that the
multi-path cyclic refinement efficiently resolves the inter-path coupling.
The NMSE drops sharply during the initial rounds and reaches a highly
stable, converged state typically within $R_{c}=3$ to $5$ rounds.
Furthermore, the results demonstrate that increasing $R_{s}$ enhances
the local precision of individual paths, enabling the cyclic procedure
to reach a lower error floor even faster. Based on this
rapid empirical convergence, we set $R_{s}=5$ and $R_{c}=5$ for
all subsequent simulations.}

We compare the proposed VNNCE algorithm with several existing algorithms.
The oracle LS method is considered as the NMSE performance bound,
assuming perfect knowledge of the true angles and distances. The far-field
SWOMP \cite{rodriguez2018frequency} and near-field P-SOMP \cite{cui2022channel}
are two on-grid algorithms, while the P-SIGW \cite{cui2022channel}
and N-NOMP \cite{lu2022near} are two near-field off-grid channel
estimation algorithms. \textcolor{black}{To ensure a fair evaluation,
the codebook generation strategies and internal optimization parameters
for all aforementioned baseline algorithms are strictly configured
according to the optimal settings recommended in their respective
original papers.} \textcolor{black}{As shown in Fig. \ref{fig:NMSE},
the proposed VNNCE algorithm significantly outperforms all the baselines
in terms of NMSE performance at all considered SNR. We can observe
that in the high SNR regimes, the performance of baselines no longer
continues to linearly decrease as SNR increases, while our proposal
can converge to }the NMSE performance bound.\textcolor{black}{{} This
implies that our algorithm, compared to the baselines, can converge
to the true solution in the high SNR regimes, thus demonstrating the
effectiveness of our designed near-field codebook and algorithm.}
\textcolor{black}{To quantitatively evaluate the computational complexity,
we measured the average CPU execution time per channel realization
for the evaluated near-field algorithms over $5000$ Monte Carlo trials,
as summarized in Table \ref{tab:RimTime}. The results demonstrate
that the proposed VNNCE algorithm maintains a highly competitive execution
time of $0.0146$ seconds. It is nearly an order of magnitude faster
than the existing near-field off-grid algorithms (P-SIGW and N-NOMP)
and operates on the same order of magnitude as the basic on-grid P-SOMP
method. This quantitative evidence is consistent with our theoretical
complexity analysis in Table \ref{tab:CCC}, confirming that the VNNCE
algorithm successfully approaches the CRLB while significantly reducing
the computational burden.}

To provide a more comprehensive evaluation of the parameter estimation
performance of the VNNCE algorithm and the cooperative localization
performance of the GFCL algorithm, we conducted a simulation using
a practical near-field 4-BS UM-MIMO system. In this system, we assume
the free space path loss model to generate the LoS channel. The channel
gain of the LoS path is given by $g_{\textrm{LoS}}=\frac{\lambda}{4\pi r_{\textrm{LoS}}}.$
Additionally, we generate an NLoS channel with random gain, angle,
and distance. The gain of the NLoS path is less than one-third of
the LoS path gain. We have included the additional key simulation
parameters in Table \ref{tab:KSPA}, while maintaining the other parameters
as listed in Table \ref{tab:KSP}. %
{} Due to the varying distances between each BS and the user, the received
SNR at each BS differs, even for the same pilot signal. 

\textcolor{black}{Fig. \ref{fig: angle-distance-Results}(a) and Fig.
\ref{fig: angle-distance-Results}(b) illustrate the }\textcolor{black}{Root
Mean Square Error (RMSE)}\textcolor{black}{{} of the propagation parameter
estimates of the LoS path, which is given by }$\mathrm{RMSE}=\sqrt{\mathrm{E}_{\hat{x}}\left\{ (x-\hat{x})^{2}\right\} }$\textcolor{black}{,
specifically angle estimate and distance estimate. The proposed VNNCE
algorithm is independently performed at each BS, while the CRLB is
derived for each BS individually. In Fig. \ref{fig: angle-distance-Results}(a),
we observe that the angle estimate achieved by the proposed VNNCE
algorithm closely approaches the level of the CRLB across all SNR
regimes. As discussed in }Sec. \ref{sec:Codebook-Design}, the angle
parameter holds greater significance in the cost function, which results
in the VNNCE algorithm attaining higher angular resolution compared
to distance resolution. This conclusion is further supported by the
results presented in \textcolor{black}{Fig. \ref{fig: angle-distance-Results}}(b).
We observe that the performance of the distance estimate closely aligns
with that of the angle estimate in high SNR regimes. However, when
the SNR is not sufficiently high, the distance estimate performance
still exhibits a gap from the theoretical estimation limit. \textcolor{black}{From
a statistical estimation perspective, this gap is theoretically anticipated:
in highly non-linear near-field spherical wave models, practical estimators
inevitably exhibit slight bias in low-SNR regimes. Nevertheless, because
the VNNCE algorithm operates over a continuous parameter space and
employs cyclic refinement to effectively decouple multi-path interference,
it acts as an asymptotically unbiased estimator \cite{kay1993fundamentals}.
Consequently, as the SNR increases, this bias diminishes, allowing
the estimation performance to converge closely to the joint CRLB.}
One noteworthy observation is that, despite BS 3 being closer to the
user compared to BS 2, implying a higher received signal SNR at BS
3, the estimation performance of both the angle and distance at BS
3 is inferior to that of BS 2. This discrepancy can be attributed
to the uniform sampling of our codebook for $\cos\theta$, and the
optimization algorithm's estimation of $\cos\theta$. As a result,
the VNNCE algorithm achieves higher angular resolution near $\theta=\pi/2$
\textcolor{black}{compared to other regions. The degradation in distance
estimation performance primarily stems from the dependence on the
angle estimation results, as evident in our codebook generation approach
outlined in }Sec. \ref{sec:Codebook-Design}. 

Fig. \ref{fig: soft-results}(a) displays the RMSE of localization
performance using different collaborative approaches. The result of
\textcolor{black}{single BS $i$ indicates localization using only BS $i$,} thus the performance of independent localization should
be equivalent to the result of distance estimation. \textcolor{black}{Most existing parameter-estimation algorithms estimate only} 
the values of propagation parameters. Therefore, the conventional
strategy for cooperative localization involves directly averaging
the user positions obtained from different BSs. \textcolor{black}{We
intentionally adopt this fundamental hard-decision strategy as our
baseline to clearly isolate and demonstrate the performance gains
explicitly brought by the proposed soft-information (confidence level)
fusion mechanism.} However, even with our proposed VNNCE algorithm,
which is capable of achieving the CRLB, the performance of the conventional
strategy may not reach the level of the best single BS localization.
From the comparison presented in Fig. \ref{fig: soft-results}(a),
we can observe that our \textcolor{black}{VNNCE-aided GFCL} algorithm has substantially
enhanced the localization accuracy from the meter-level to the millimeter-level.
This demonstrates the effectiveness of our approach in achieving highly
accurate user localization. Furthermore, the results underscore the
\textcolor{black}{value of uncertainty-aware soft information} in our high-precision localization
system. 

Fig. \ref{fig: soft-results}(b) compares the NMSE of the channel
estimation performance in Steps 1 and 3 in the joint architecture,
considering each BS individually. The results clearly demonstrate
that the channel refinement step leads to a notable improvement of
approximately 1.4 dB in the channel estimation accuracy for BS 1.
By referring to Fig. \ref{fig:NMSE}, we can observe that this improvement
enables the channel estimation results to closely approach the theoretical
limit, i.e., the performance level of the oracle LS method. This signifies
the effectiveness of the channel refinement process in enhancing the
accuracy and quality of the channel estimation results, thereby bringing
them closer to the theoretical limits of the system.

\section{Conclusion \label{sec:Conclusions}}

In this paper, we propose a unified framework for joint channel estimation
and cooperative localization in near-field UM-MIMO systems. The proposed
VNNCE algorithm addresses the limitations of existing off-grid channel
estimation methods by providing a low-complexity, tuning-free solution
with \textcolor{black}{local} convergence guarantees, achieving channel
estimation accuracy approaching the CRLB. \textcolor{black}{The VNNCE algorithm also provides soft estimates with associated confidence information, which are exploited by the GFCL algorithm for channel-estimation-aided cooperative localization, thereby improving localization accuracy and robustness.} Conversely, the high-accuracy user position
estimate obtained by the GFCL algorithm is fed back to refine channel
estimation, realizing cooperative localization\textendash enhanced
channel estimation. This bidirectional integration yields consistent
performance gains in both tasks, demonstrating the effectiveness of
joint design in near-field UM-MIMO systems. 

\textcolor{black}{While the proposed VNNCE-GFCL framework demonstrates
performance close to the CRLB in the considered simulations, we acknowledge that
the current performance evaluations are based on a relatively idealized
baseline model. In practical deployment scenarios, near-field localization
systems face severe impairments such as LoS blockage, dense NLoS interference,
closely spaced multipath clusters, and array geometry mismatches.
Addressing these complex environmental factors requires advanced robust
statistics and outlier rejection mechanisms. Additionally, integrating
more advanced standalone near-field localization algorithms into our
soft-information fusion framework to further elevate the absolute
positioning accuracy represents another important direction for our
future work.}

\appendix{}

\section*{\textcolor{black}{Analysis of the Hessian Negative Definiteness\label{sec:Appendix}}}

\textcolor{black}{In this appendix, we rigorously prove that using the inverse Hessian matrix to obtain the covariance (soft information)
is mathematically valid under the proposed alternating optimization
framework. Specifically, we demonstrate that if the 2D concentrated
objective function with respect to the spatial parameters is locally
concave at a stationary point, the full four-dimensional (4D) Hessian
matrix of the joint parameter space is strictly negative definite,
which guarantees a valid positive definite covariance matrix.}

\textcolor{black}{Let $\mathbf{y}\in\mathbb{C}^{M}$ denote the received
signal vector and $\mathbf{b}(\boldsymbol{\xi})\in\mathbb{C}^{M}$
denote the near-field steering vector, where the spatial parameter
vector is defined as $\boldsymbol{\xi}=[\theta,r]^{T}$. The steering
vector satisfies the constant modulus property, yielding $\|\mathbf{b}(\boldsymbol{\xi})\|^{2}=M$.
To avoid the singularity of the Hessian matrix at $g=0$ caused by
the polar coordinate representation $ge^{j\phi}$, we redefine the
complex channel gain in its Cartesian form $\alpha=ge^{j\phi}=\alpha_{R}+j\alpha_{I}$.
Letting the linear gain parameter vector be denoted by $\boldsymbol{\beta}=[\alpha_{R},\alpha_{I}]^{T}$,
the full unknown parameter vector is thus defined as $[\boldsymbol{\xi}^{T},\boldsymbol{\beta}^{T}]^{T}$
representing the 4D search space. The AWGN observation model is expressed
as $\mathbf{y}=\alpha\mathbf{b}(\boldsymbol{\xi})+\mathbf{n}$, where
$\mathbf{n}\sim\mathcal{CN}(\mathbf{0}_{M},\sigma^{2}\mathbf{I}_{M})$
models the noise component.}

\textcolor{black}{Under the AWGN assumption, maximizing the log-likelihood
is mathematically equivalent to maximizing the negative squared Euclidean
distance. By dropping the constant term $-\|\mathbf{y}\|^{2}/\sigma^{2}$
and defining the scalar spatial projection function as $c(\boldsymbol{\xi})=\mathbf{b}(\boldsymbol{\xi})^{H}\mathbf{y}$,
the objective simplifies to the following formulation, given by}

\textcolor{black}{
\begin{equation}
f(\boldsymbol{\xi},\boldsymbol{\beta})=\frac{1}{\sigma^{2}}\left[2\Re\{\alpha^{*}c(\boldsymbol{\xi})\}-M|\alpha|^{2}\right].\label{eq:Afunc}
\end{equation}
For any fixed spatial parameter $\boldsymbol{\xi}$, the objective
function $f(\boldsymbol{\xi},\boldsymbol{\beta})$ is a strictly concave
quadratic function with respect to the linear gain $\boldsymbol{\beta}$.
By taking the derivative with respect to $\alpha^{*}$ and equating
it to zero, the conditionally global optimal complex gain is obtained
as 
\begin{equation}
\hat{\alpha}(\boldsymbol{\xi})=\frac{c(\boldsymbol{\xi})}{M}=\frac{\mathbf{b}(\boldsymbol{\xi})^{H}\mathbf{y}}{M}.\label{eq:Aalpha}
\end{equation}
Substituting $\hat{\alpha}(\boldsymbol{\xi})$ back into \eqref{eq:Afunc}
yields the 2D concentrated objective function $F(\boldsymbol{\xi})$,
which depends only on the spatial parameters and is given by 
\begin{equation}
F(\boldsymbol{\xi})=f(\boldsymbol{\xi},\hat{\boldsymbol{\beta}}(\boldsymbol{\xi}))=\frac{|\mathbf{b}(\boldsymbol{\xi})^{H}\mathbf{y}|^{2}}{M\sigma^{2}}.\label{eq:AfuncO}
\end{equation}
Note that $F(\boldsymbol{\xi})$ is directly proportional to the cost
function $G_{y}(\theta,r)$ \eqref{eq:costFunc} proposed in our detection
stage, which theoretically verifies the equivalence of the optimization
objectives. }

\textcolor{black}{Using the total derivative chain rule, the gradient
of the concentrated objective $F(\boldsymbol{\xi})$ is derived as
\begin{equation}
\nabla_{\boldsymbol{\xi}}F(\boldsymbol{\xi})=\nabla_{\boldsymbol{\xi}}f(\boldsymbol{\xi},\hat{\boldsymbol{\beta}}(\boldsymbol{\xi}))+\left(\frac{\partial\hat{\boldsymbol{\beta}}(\boldsymbol{\xi})}{\partial\boldsymbol{\xi}}\right)^{T}\nabla_{\boldsymbol{\beta}}f(\boldsymbol{\xi},\hat{\boldsymbol{\beta}}(\boldsymbol{\xi})).\label{eq:A1stD}
\end{equation}
Since $\hat{\boldsymbol{\beta}}(\boldsymbol{\xi})$ is the exact optimal
solution for a given $\boldsymbol{\xi}$, the first-order optimality
condition dictates that $\nabla_{\boldsymbol{\beta}}f(\boldsymbol{\xi},\hat{\boldsymbol{\beta}}(\boldsymbol{\xi}))=\mathbf{0}$.
Consequently, the second term vanishes to produce 
\begin{equation}
\nabla_{\boldsymbol{\xi}}F(\boldsymbol{\xi})=\nabla_{\boldsymbol{\xi}}f(\boldsymbol{\xi},\hat{\boldsymbol{\beta}}(\boldsymbol{\xi})).\label{eq:A1stD2}
\end{equation}
This mathematical relationship proves that any stationary point $\nabla_{\boldsymbol{\xi}}F(\boldsymbol{\xi})=\mathbf{0}$
in the 2D concentrated space corresponds to a stationary
point in the full 4D parameter space.}

\textcolor{black}{The full 4D Hessian matrix $\mathbf{H}$ of $f(\boldsymbol{\xi},\boldsymbol{\beta})$
can be expressed in a partitioned block structure formulated as 
\begin{equation}
\mathbf{H}=\begin{bmatrix}\mathbf{H}_{\boldsymbol{\xi}\boldsymbol{\xi}} & \mathbf{H}_{\boldsymbol{\xi}\boldsymbol{\beta}}\\
\mathbf{H}_{\boldsymbol{\beta}\boldsymbol{\xi}} & \mathbf{H}_{\boldsymbol{\beta}\boldsymbol{\beta}}
\end{bmatrix}.
\end{equation}
Because $f(\boldsymbol{\xi},\boldsymbol{\beta})$ in \eqref{eq:Afunc}
is quadratic with respect to the linear parameters $\boldsymbol{\beta}=[\alpha_{R},\alpha_{I}]^{T}$,
its corresponding sub-Hessian $\mathbf{H}_{\boldsymbol{\beta}\boldsymbol{\beta}}$
is invariant to the parameters and unconditionally strictly negative
definite. This property guarantees 
\begin{equation}
\mathbf{H}_{\boldsymbol{\beta}\boldsymbol{\beta}}=-\frac{2M}{\sigma^{2}}\mathbf{I}_{2}\prec0.\label{eq:AHessianbb}
\end{equation}
To find the exact relationship between the 2D Hessian $\nabla_{\boldsymbol{\xi}}^{2}F(\boldsymbol{\xi})$
and the full 4D Hessian $\mathbf{H}$, we take the total derivative
of the optimality condition $\nabla_{\boldsymbol{\beta}}f(\boldsymbol{\xi},\hat{\boldsymbol{\beta}}(\boldsymbol{\xi}))=\mathbf{0}$
with respect to $\boldsymbol{\xi}$, which yields 
\begin{equation}
\frac{\partial\hat{\boldsymbol{\beta}}}{\partial\boldsymbol{\xi}}=-\mathbf{H}_{\boldsymbol{\beta}\boldsymbol{\beta}}^{-1}\mathbf{H}_{\boldsymbol{\beta}\boldsymbol{\xi}}.
\end{equation}
Taking the derivative of \eqref{eq:A1stD2} with respect to $\boldsymbol{\xi}$
produces the expanded Hessian term 
\begin{equation}
\nabla_{\boldsymbol{\xi}}^{2}F(\boldsymbol{\xi})=\mathbf{H}_{\boldsymbol{\xi}\boldsymbol{\xi}}+\mathbf{H}_{\boldsymbol{\xi}\boldsymbol{\beta}}\frac{\partial\hat{\boldsymbol{\beta}}}{\partial\boldsymbol{\xi}}=\mathbf{H}_{\boldsymbol{\xi}\boldsymbol{\xi}}-\mathbf{H}_{\boldsymbol{\xi}\boldsymbol{\beta}}\mathbf{H}_{\boldsymbol{\beta}\boldsymbol{\beta}}^{-1}\mathbf{H}_{\boldsymbol{\beta}\boldsymbol{\xi}}.\label{eq:AHessian}
\end{equation}
To interpret this result, we recall that for the partitioned matrix
$\mathbf{H}$ defined previously, its Schur Complement $\mathbf{S}$
with respect to the block $\mathbf{H}_{\boldsymbol{\beta}\boldsymbol{\beta}}$
is formally defined as 
\begin{equation}
\mathbf{S}=\mathbf{H}_{\boldsymbol{\xi}\boldsymbol{\xi}}-\mathbf{H}_{\boldsymbol{\xi}\boldsymbol{\beta}}\mathbf{H}_{\boldsymbol{\beta}\boldsymbol{\beta}}^{-1}\mathbf{H}_{\boldsymbol{\beta}\boldsymbol{\xi}}.\label{eq:ASchur}
\end{equation}
Comparing \eqref{eq:AHessian} and \eqref{eq:ASchur} reveals a fundamental
mathematical property of the variable projection method \cite{Golub1973VarPro}
where the Hessian of the 2D concentrated objective function $\nabla_{\boldsymbol{\xi}}^{2}F(\boldsymbol{\xi})$
is mathematically identical to the Schur Complement $\mathbf{S}$.}

\textcolor{black}{According to the classical Schur complement condition
for matrix definiteness \cite{Horn2012Matrix}, a partitioned symmetric
matrix $\mathbf{H}$ is strictly negative definite if and only if
its diagonal block $\mathbf{H}_{\boldsymbol{\beta}\boldsymbol{\beta}}$
and its corresponding Schur complement $\mathbf{S}$ are both strictly
negative definite. Given that the sub-Hessian satisfies $\mathbf{H}_{\boldsymbol{\beta}\boldsymbol{\beta}}\prec0$
unconditionally as derived in \eqref{eq:AHessianbb}, the negative
definiteness of the full 4D Hessian $\mathbf{H}$ is dictated entirely
by $\mathbf{S}$. We formally establish this equivalency as
\begin{equation}
\mathbf{H}\prec0\iff\mathbf{S}=\nabla_{\boldsymbol{\xi}}^{2}F(\boldsymbol{\xi})\prec0.\label{eq:Aequ}
\end{equation}
}

\textcolor{black}{The proof provided in \eqref{eq:Aequ} confirms that
if the optimization algorithm successfully converges to a stationary
point where the 2D spatial Hessian $\nabla_{\boldsymbol{\xi}}^{2}F(\boldsymbol{\xi})$
is negative definite, the full 4D Hessian matrix $\mathbf{H}$ at
the reconstructed joint parameter coordinates is mathematically guaranteed
to be strictly negative definite. In our framework, this local concavity
requirement is explicitly enforced by the algorithmic checks in Algorithm
\ref{alg:NewtonUpdate}, ensuring that extracting the covariance matrix
via the inverse of the negative full Hessian inherently yields a valid
and positive definite covariance matrix.}

\bibliographystyle{IEEEtran}
\bibliography{reference}

\end{document}